\documentclass[preprint,aps,prd,showpacs,superscriptaddress,nofootinbib,tightenlines]{revtex4}

\usepackage{amsfonts}
\usepackage{mathrsfs}
\usepackage{graphicx}
\usepackage{amsmath}
\usepackage{amssymb}
\usepackage{bm}
\usepackage{bbm}
\usepackage{epsfig}
\usepackage{multirow}
\usepackage{float}
\usepackage{array}
\makeatletter
\def\hlinew#1{%
  \noalign{\ifnum0=`}\fi\hrule \@height #1 \futurelet
   \reserved@a\@xhline}
\usepackage{array}
\newcommand{\PreserveBackslash}[1]{\let\temp=\\#1\let\\=\temp}
\newcolumntype{C}[1]{>{\PreserveBackslash\centering}p{#1}}
\newcolumntype{R}[1]{>{\PreserveBackslash\raggedleft}p{#1}}
\newcolumntype{L}[1]{>{\PreserveBackslash\raggedright}p{#1}}

\def\bfalpha{\mbox{\boldmath $\alpha$}}
\def\bfpsi{\mbox{\boldmath $\psi$}}
\def\bfchi{\mbox{\boldmath $\chi$}}

\def\OMIT#1{}

\newcommand{\nn}{\nonumber}

\newcommand{\beq}{\begin{equation}}
\newcommand{\eeq}{\end{equation}}
\newcommand{\bqa}{\begin{eqnarray}}
\newcommand{\eqa}{\end{eqnarray}}

\begin{document}

\title{\mbox{}\\[11pt]
$\bm{\mathcal O}\bm{(}\bfalpha_{\bm s}\bm{)}$ corrections to
$\bm{J}\bm{/}\bfpsi \bm{+}\bfchi_{\bm{c}\bm{J}}$ production at
$\bm{B}$ factories}


\renewcommand{\theequation}{\thesection.\arabic{equation}}
\renewcommand{\thesection}{\arabic{section}}
\renewcommand{\thetable}{\arabic{table}}
\renewcommand{\thefigure}{\arabic{figure}}

\author{Hai-Rong Dong\footnote{E-mail: donghr@ihep.ac.cn}}
\affiliation{Institute of High Energy Physics, Chinese Academy of
Sciences, Beijing 100049, China\vspace{0.2cm}}

\author{Feng Feng\footnote{E-mail: fengf@ihep.ac.cn}}
\affiliation{Theoretical Physics Center for Science Facilities,
Chinese Academy of Sciences, Beijing 100049, China\vspace{0.2cm}}
\affiliation{Institute of High Energy Physics, Chinese Academy of
Sciences, Beijing 100049, China\vspace{0.2cm}}

\author{Yu Jia\footnote{E-mail: jiay@ihep.ac.cn}}
\affiliation{Maryland Center for Fundamental Physics, Department of
Physics, University of Maryland, College Park, MD 20742,
USA\vspace{0.2cm}}
\affiliation{Institute of High Energy Physics,
Chinese Academy of Sciences, Beijing 100049, China\vspace{0.2cm}}
\affiliation{Theoretical Physics Center for Science Facilities,
Chinese Academy of Sciences, Beijing 100049, China\vspace{0.2cm}}

\date{\today}
\begin{abstract}

We investigate the ${\cal O}(\alpha_s)$ corrections to $e^+e^-\to
J/\psi(\psi')+\chi_{cJ}$ ($J=0,1,2$) in the NRQCD factorization
approach. These next-to-leading order (NLO) corrections are
calculated at the level of helicity amplitude. We have made a
detailed analysis for both polarized and unpolarized cross sections,
and compared our predictions with the measurements at the $B$
factories. We also derive the asymptotic expressions for each of the
NLO helicity amplitudes, and confirm the earlier speculation that at
NLO in $\alpha_s$, the double logarithm of type $\ln^2 (s/m_c^2)$
appearing in the NRQCD short-distance coefficient is always
associated with the helicity-suppressed channels.

\end{abstract}

\pacs{\it 12.38.Bx, 13.66.Bc, 13.88.+e, 14.40.Pq}


\maketitle

\section{Introduction}
\label{Introduction}
\setcounter{equation}{0}

The studies of hard exclusive reactions have historically played an
important role in the development of perturbative Quantum
Chromodynamics (pQCD). The standard theoretical tool is the
light-cone (collinear factorization) approach, which is based on the
expansion in powers of $1/Q$, where $Q^2$ characterizes the hard
momentum transfer~\cite{Lepage:1980fj,Chernyak:1983ej}. The classic
applications of light-cone factorization are exemplified by the
$\pi\!-\!\gamma$ transition form factor, $\pi$ electromagnetic (EM)
form factor~\cite{Lepage:1980fj,Chernyak:1983ej}, and the $B$ meson
exclusive decays~\cite{Beneke:1999br,Beneke:2000ry}, to which a vast
amount of literature has been devoted.

Recent advancement of the high-luminosity high-energy collider
facilities renders the study of hard exclusive processes involving
heavy quarkonium a fertile new research frontier. Perhaps a great
amount of interest toward this topic was triggered by the
observation of several double-charmonium production processes in two
$B$ factories several years ago~\cite{Abe:2002rb,Aubert:2005tj}.

In addition to the light-cone approach, there also exists another
relatively new factorization approach, the {\it NRQCD
factorization}~\cite{Bodwin:1994jh}, which is tailor-made to tackle
quarkonium production and decay processes. Explicitly exploiting the
nonrelativistic nature of quarkonium, the NRQCD factorization
approach allows one to express the amplitude of an exclusive
quarkonium production reaction in terms of an infinite sum of
products of short-distance coefficients and the vacuum-to-quarkonium
NRQCD matrix elements, whose importance is organized by the powers
of $v$, the typical velocity of a heavy quark inside the quarkonium.

Perhaps the most famous example of double charmonium production is
$e^+e^-\to J/\psi+\eta_c$, with the interest originally spurred by
the alarming discrepancy between the \textsc{Belle}
measurement~\cite{Abe:2002rb} and the leading-order (LO) NRQCD
predictions~\cite{Braaten:2002fi,Liu:2002wq,Hagiwara:2003cw}. In the
following years, this process has been intensively studied in both
of the NRQCD and light-cone
frameworks~\cite{Braaten:2002fi,Liu:2002wq,Hagiwara:2003cw,
Zhang:2005cha,Gong:2007db,Ma:2004qf,Bondar:2004sv,Braguta:2008tg}.

One of the important steps toward alleviating the discrepancy
between data and theory for $e^+e^-\to J/\psi+\eta_c$ is the
discovery of significant and positive NLO perturbative
corrections~\cite{Zhang:2005cha,Gong:2007db}. This NLO calculation
was performed in NRQCD factorization. In contrast, due to some
long-standing theoretical difficulty inherent to this
helicity-suppressed process, by far no one has successfully worked
out the NLO correction to this process in the light-cone approach.

Besides $e^+e^-\to J/\psi+\eta_c$, two $B$ factories have also
measured several other double-charmonium production
processes~\cite{Abe:2002rb,Aubert:2005tj}, notably $e^+e^-\to
J/\psi+\chi_{cJ}$ ($J=0,1,2$). Recently in Ref.~\cite{Zhang:2008gp},
the NLO perturbative correction was performed for $e^+e^-\to
J/\psi+\chi_{c0}$ at LO in $v$ and a rather large K factor was
reported to bring the theory prediction in better agreement with the
measurement. Nevertheless, at this stage, both $B$ factory
experiments have not clearly observed any $J/\psi+\chi_{c1,2}$
events yet, even with latest data set~\cite{:2009nj}, and only an
upper bound for the production cross sections is placed.

The purpose of this work is to carry out a comprehensive
next-to-leading order (NLO) perturbative analysis to $e^+e^-\to
J/\psi(\psi')+\chi_{cJ}$ ($J=0,1,2$) at the lowest order in $v$. We
will work out the NLO corrections to each of the helicity
amplitudes, so that we can address their impact on both the
polarized and unpolarized cross sections.

We find a large positive NLO perturbative correction to $e^+e^-\to
J/\psi+\chi_{c0}$, which is helpful to bring the predicted cross
section closer to the $B$ factory measurements. However, our NLO
predictions to $\psi'+\chi_{c0}$ cross section seems still
significantly below the central value of the \textsc{Belle}
measurement.

On the other hand, the impact of NLO corrections to $e^+e^-\to
J/\psi+\chi_{c1,2}$ seems to be rather modest, even with their signs
uncertain. Our predicted cross sections for these processes are
about 5 or 6 times smaller than that for $e^+e^-\to
J/\psi+\chi_{c0}$. Hopefully the future Super $B$ factory, with much
higher luminosity, will eventually observe these two channels.

Our studies of polarized cross sections reveal that the bulk of the
total cross section comes from the $(0,\pm 1)$ helicity states for
$e^+e^-\to J/\psi+\chi_{c1}$, and from the $(0,0)$ and $(\pm 1,0)$
helicity states for $e^+e^-\to J/\psi+\chi_{c2}$. It will be
interesting for the future Super $B$ experiments to test these
polarization patterns.

From the theoretical perspective, we have also presented the
analytic asymptotic expressions of all the ten NLO helicity
amplitudes for the $e^+e^-\to J/\psi+\chi_{cJ}$  ($J=0,1,2$)
processes. The logarithmical scaling-violation pattern is recognized
in these NLO asymptotic expressions, which supports for the
speculation made in Ref.~\cite{Jia:2010fw}: The hard exclusive
processes involving double-quarkonium at leading twist can only host
the single collinear logarithm $\ln{s/m_c^2}$ at one-loop, while
those beginning with the higher twist contributions are always
plagued with double logarithms of form $\ln^2{s/m_c^2}$. It would be
theoretically interesting to reproduce these asymptotic expressions
by using the light-cone approach.

The rest of the paper is organized as follows.
In Section~\ref{pol:cross:sect:hsr}, we present a concise review on
the helicity amplitude formalism and helicity selection rule, taking
the process $e^- e^+ \to J/\psi + \chi_{cJ}$ ($J=0,1,2$) as the
example.
In Section~\ref{LO:result}, we rederive the tree-level predictions
in the NRQCD factorization approach, and present the LO expressions
for all the involved helicity amplitudes.
In Section~\ref{NLO:result}, we first elaborate on some key
technical issues about the NLO perturbative calculations, then
present the asymptotic expressions for the NLO corrections for all
the encountered helicity amplitudes. The pattern of the logarithmic
scaling violation is recognized, and their implication with the
light-cone approach is discussed.
We devote Section~\ref{phenomenology} to explore the
phenomenological impact of our NLO predictions on the $B$ factory
measurements. A comprehensive numerical study is performed for both
unpolarized and polarized cross sections of the $e^- e^+ \to J/\psi
+ \chi_{cJ}$ ($J=0,1,2$) processes.
Finally, we summarize in Section~\ref{phenomenology}.

\section{Polarized cross sections and helicity selection rule}
\label{pol:cross:sect:hsr}
\setcounter{equation}{0}

It is often desirable to know the polarization information for an
exclusive reaction, especially for the double-charmonium production
process considered in this work. To fulfill this goal, one can
employ the {\it helicity amplitude formalism}~\cite{Jacob:1959at},
which appears to be a quite convenient and advantageous tool both
experimentally and theoretically.

Suppose the $e^-$ and $e^+$ beams are aligned along the $\hat{z}$
direction, bearing the invariant mass of $\sqrt{s}$. We will choose
to work in their center-of-mass frame. Let $\theta$ denote the angle
between the moving directions of $J/\psi$ and $e^-$, and $|{\bf P}|$
signify the magnitude of the momentum carried by the $J/\psi$
($\chi_{cJ}$) [Note that $|{\bf P}|=\lambda^{1\over
2}(s,M^2_{J/\psi},M^2_{\chi_{cJ}})/(2\sqrt{s})$, where
$\lambda(x,y,z)=x^2+y^2+z^2-2xy-2yz-2zx$]. Assume the outgoing
$J/\psi$ and $\chi_{cJ}$ carry definite helicities of $\lambda_1$,
$\lambda_2$, respectively. First let us imagine the process of a
virtual photon decay into $J/\psi$ and $\chi_{cJ}$. The differential
decay rate can be expressed as~\cite{Jacob:1959at,Haber:1994pe}
\bqa
{d\Gamma [\gamma^*(S_z)\to
J/\psi(\lambda_1)+\chi_{cJ}(\lambda_2)]\over d\cos \theta} &=&
{|{\bf P}|\over 16\pi s} \left| d^1_{S_z,\lambda}(\theta) \right|^2
\left|{\mathcal A}^J_{\lambda_1,\lambda_2}\right|^2,
\label{virtual:photon:diff:decay:rate}
\eqa
where  $\lambda=\lambda_1-\lambda_2$, $S_z$ is the spin projection
of the virtual photon along the $\hat{z}$ axis. ${\mathcal
A}^J_{\lambda_1,\lambda_2}$ is the desired helicity amplitude, which
encodes all the nontrivial dynamics. The angular distribution is
fully dictated by the quantum numbers $S_z$ and $\lambda$ through
the Wigner rotation matrix $d^j_{m,m'}(\theta)$. Note angular
momentum conservation constrains that $|\lambda|\le 1$.

The number of independent helicity amplitudes can be greatly
reduced, thanks to the parity invariance:
\bqa
{\mathcal A}^J_{\lambda_1,\lambda_2}=(-1)^J {\mathcal
A}^J_{-\lambda_1,-\lambda_2}.
\label{parity:trans:hel:ampl}
\eqa
Consequently, the two helicity amplitudes related by simultaneously
flipping the helicities of $J/\psi$ and $\chi_{cJ}$ bear the equal
magnitude.

It is straightforward to convert the differential decay rate
(\ref{virtual:photon:diff:decay:rate}) to the corresponding
production cross section in $e^+e^-$ annihilation:
\bqa
& & {d\sigma [e^+e^-\to J/\psi(\lambda_1)+\chi_{cJ}(\lambda_2)]\over
d\cos \theta} = {2\pi\alpha \over s^{3/2}} \sum_{S_z=\pm 1} {d\Gamma
[\gamma^*(S_z)\to J/\psi(\lambda_1)+\chi_{cJ}(\lambda_2)]\over d\cos
\theta}
 \nn\\
 & & = {\alpha \over 8 s^2}
\left({|{\bf P}|\over \sqrt{s}}\right) |{\mathcal
A}^J_{\lambda_1,\lambda_2}|^2 \times \Bigg\{
\begin{array}{c}
{1+\cos^2\theta\over 2},\qquad\qquad(\lambda={\pm 1})
\\
\sin^2\theta,\qquad\qquad (\lambda=0)
\end{array}
\label{polar:diff:cross:section}
\eqa
where the polarizations of $e^-$ and $e^+$ have been averaged over.
Note that we only need to sum over two {\it transverse}
polarizations of the virtual photon, as guaranteed by the helicity
conservation in QED for a massless electron. This selective
summation is the cause for the anisotropic angular distribution
patterns in (\ref{polar:diff:cross:section}).

It is now straightforward to acquire the unpolarized cross sections.
Integrating (\ref{polar:diff:cross:section}) over the polar angle
$\theta$ and including all the allowed helicity states, one finally
arrives at:
\begin{subequations}
\bqa
& & \sigma[J/\psi + \chi_{c0}] = {\alpha\over 6 s^2} \left({|{\bf
P}|\over \sqrt{s}}\right) \left( \left|{\mathcal
A}^0_{0,0}\right|^2+2 \left|{\mathcal A}^0_{1,0}\right|^2\right),
\\
& & \sigma[J/\psi + \chi_{c1}] = {\alpha\over 6 s^2} \left({|{\bf
P}|\over \sqrt{s}}\right) \left(2 \left|{\mathcal
A}^1_{1,0}\right|^2+ 2 \left|{\mathcal A}^1_{0,1}\right|^2 +  2
\left|{\mathcal A}^1_{1,1}\right|^2 \right),
\\
& & \sigma[J/\psi + \chi_{c2}] = {\alpha\over 6 s^2} \left({|{\bf
P}|\over \sqrt{s}}\right) \left(\left|{\mathcal A}^2_{0,0}\right|^2+
2 \left|{\mathcal A}^2_{1,0}\right|^2 +  2 \left|{\mathcal
A}^2_{0,1}\right|^2 + 2 \left|{\mathcal A}^2_{1,1}\right|^2 +  2
\left | {\mathcal A}^2_{1,2}\right|^2 \right).
\eqa
\label{unpol:cross:section:Jpsi:chicJ}
\end{subequations}
There are 2, 3 and 5 independent helicity amplitudes for $\gamma^*
\to J/\psi+\chi_{cJ}$ ($J=0,1,2$), respectively, as required by the
angular momentum conservation. Note that parity transformation
property (\ref{parity:trans:hel:ampl}) forbids the appearance of the
${\mathcal A}^{1}_{0,0}$ amplitude. In equation
(\ref{unpol:cross:section:Jpsi:chicJ}), we include a factor of 2 to
account for the parity-doublet contribution.

Aside from its great technical usefulness, the helicity amplitude
formalism can also shed important light on the dynamics underlying
exclusive reactions. In particular, each helicity amplitude
possesses a definite power-law scaling in $1/s$, controlled by the
{\it helicity selection rule}~\cite{Brodsky:1981kj}. At
asymptotically large $\sqrt{s}$, the polarized cross section for our
double-charmonium production process scales
as~\cite{Braaten:2002fi}:
\bqa
\sigma[J/\psi(\lambda_1) + \chi_{cJ}(\lambda_2)] & \sim & \alpha^2
v^8 \left(m_c^2\over s \right)^{2+|\lambda_1+\lambda_2|},
\label{helicity:selection:rule}
\eqa
here $v$ denotes the characteristic velocity of charm quark inside a
charmonium.

Equation (\ref{helicity:selection:rule}), combined with angular
momentum conservation, implies that the helicity state which
exhibits the slowest asymptotic decrease, {\it i.e.} $\sigma \propto
1/s^2$, is the one that conserves the hadron helicities
$|\lambda_1+\lambda_2|=0$, or equivalently, $(\lambda_1,
\lambda_2)=(0,0)$. For each unit of the violation of this law, there
is a further suppression factor of $1/s$. At sufficiently high
energy, only the $(0,0)$ helicity state perhaps needs to be retained
for phenomenological purposes. Note that in NRQCD factorization
language, the charm quark is treated as heavy, and in fact its mass
acts as the agent of violating the hadron helicity conservation.

Once beyond the lowest order in $\alpha_s$, the power-law scaling
specified in (\ref{helicity:selection:rule}) will in general receive
mild modifications due to the $\ln s$ terms. This logarithmic
scaling violation will be examined in detail in
Section~\ref{NLO:result}.

\section{LO predictions for polarized cross sections}
\label{LO:result}
\setcounter{equation}{0}

\begin{figure}[t]
\begin{center}
\includegraphics[scale=0.65]{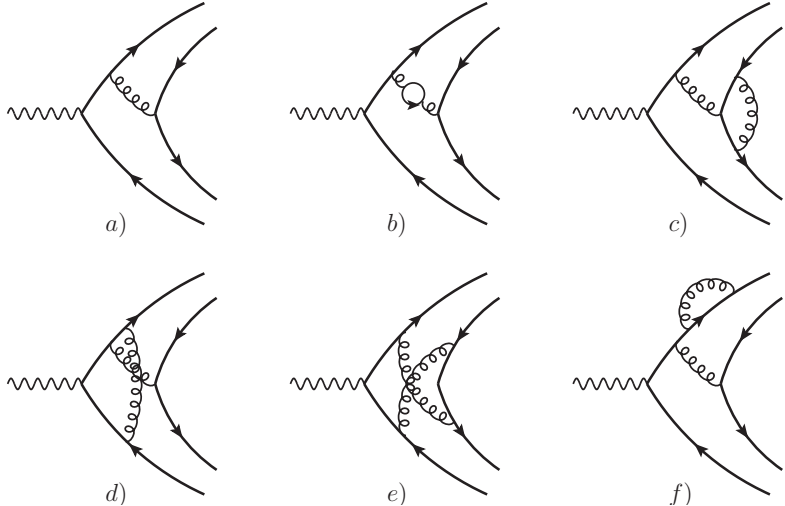}
\caption{One sample LO diagram and five sample NLO diagrams that
contribute to $\gamma^*\to J/\psi+\chi_{cJ}$.
\label{feynman:diagrams}}
\end{center}
\end{figure}

At LO in $v$ expansion in the NRQCD approach, one can factorize the
amplitude of $\gamma^*\to J/\psi+\chi_{cJ}$ as the product of the
short-distance coefficients and the nonperturbative NRQCD matrix
elements. To a good extent, these nonperturbative matrix elements
are well approximated by the phenomenological (derivative of) wave
function at the origin, $R_{J/\psi}(0)$ and $R'_{\chi_{cJ}}(0)$. At
LO in $\alpha_s$, the short-distance coefficients have been
obtained~\cite{Braaten:2002fi,Liu:2002wq} by calculating the quark
amplitude $\gamma^*\to c\bar{c} (^3S_1^{(1)})+c\bar{c}
(^3P_J^{(1)})$ using the covariant projection
technique~\cite{Kuhn:1979bb,Bodwin:2002hg}.

There are in total 4 Feynman diagrams at LO in $\alpha_s$, one of
which is depicted in Fig.~\ref{feynman:diagrams}$a)$. For
simplicity, we have neglected the QED fragmentation diagrams, while
their effect is modest. To expedite the extraction of the
corresponding helicity amplitudes, we have constructed 10 helicity
projection operators. It is convenient to parameterize the LO
helicity amplitude as
\bqa
{{\mathcal A}^J}^{(0)}_{\lambda_1,\lambda_2}  &=&  {4 e e_c \alpha_s
C_F R_{J/\psi}(0) R'_{\chi_{cJ}}(0)\over m_c^3}\,r^{{1\over
2}(1+|\lambda_1+\lambda_2|)} \,c^J_{\lambda_1,\lambda_2}(r),
\label{LO:helic:ampl:parametrization}
\eqa
where $e_c={2\over 3}$ is the electric charge of the charm quark,
and for brevity, we have introduced a dimensionless variable $r$:
\bqa
r & \equiv & {4 m_c^2 \over s}.
\eqa
To make the scaling behavior in (\ref{helicity:selection:rule})
transparent, we have explicitly stripped off a factor $r^{{1\over
2}(1+|\lambda_1+\lambda_2|)}$ in
(\ref{LO:helic:ampl:parametrization}), so that the reduced helicity
amplitude $c^J_{\lambda_1,\lambda_2}(r)$, a dimensionless function,
will start with a ${\cal O}(1)$ constant. Concretely, these
$c^J_{\lambda_1,\lambda_2}(r)$ functions are
\begin{subequations}
\bqa
& & c^0_{0,0}(r)= 1+ 10r - 12r^2\qquad c^0_{1,0}(r)= 9-14r,
\\
& & c^1_{0,1}(r)=-\sqrt{6}(2-7 r)\qquad c^1_{1,0}(r)=-\sqrt{6}
r\qquad c^1_{1,1}(r)= -2\sqrt{6}(1-3r),
\label{c:function:LO:chic1}
\\
& & c^2_{0,0}(r)=\sqrt{2}(1-2r-12r^2)\qquad
c^2_{0,1}(r)=\sqrt{6}(1-5r)\qquad c^2_{1,0}(r)= \sqrt{2}(3-11r),
\nn\\
& & c^2_{1,1}(r) = 2\sqrt{6}(1-3r)\qquad c^2_{1,2}(r)= 2\sqrt{3}.
\eqa
\end{subequations}
These 10 helicity amplitudes agree with those given in
Ref.~\cite{Braaten:2002fi}. It is interesting to note that the
tree-level $J/\psi(1)+\chi_{c1}(0)$ amplitude constitutes an
exception in that it accidently receives an extra suppression factor
than expected from the helicity selection rule.

Plugging (\ref{LO:helic:ampl:parametrization}) into
(\ref{unpol:cross:section:Jpsi:chicJ}), we obtain the LO NRQCD
predictions for the polarized cross sections:
\bqa
& & \sigma^{(0)}[J/\psi(\lambda_1) + \chi_{cJ}(\lambda_2)] = {32\pi
e_c^2 \alpha^2 C_F^2 \alpha_s^2 \over 3 s^2 m_c^6} \left({|{\bf
P}|\over \sqrt{s}}\right) R^2_{J/\psi}(0) R'^2_{\chi_{cJ}}(0)
r^{1+|\lambda_1+\lambda_2|} \left|
c^J_{\lambda_1,\lambda_2}(r)\right|^2.
\nn\\
\label{pol:cross:section:LO:alphas}
\eqa

\section{NLO perturbative corrections to polarized cross sections}
\label{NLO:result}
\setcounter{equation}{0}

In this section, we first sketch some technical issues about the NLO
perturbative calculations, then present the asymptotic expressions
of the NLO corrections for all the encountered helicity amplitudes.

\subsection{Description of the calculation}

We first employ the \textsc{Mathematica} package
\textsc{FeynArts}~\cite{Kublbeck:1990,Hahn:2000kx} to generate
Feynman diagrams and amplitudes for the parton process $\gamma^*\to
c\bar{c}(P_1) +c\bar{c}(P_2)$ to NLO in $\alpha_s$. Feynman gauge is
used throughout this calculation. In total there are 20 two-point,
20 three-point, 18 four-point, and 6 five-point one-loop diagrams,
some of which have been illustrated in Fig.~\ref{feynman:diagrams}.
We then apply the covariant projector
technique~\cite{Kuhn:1979bb,Bodwin:2002hg} to enforce that two
$c\bar{c}$ pairs form the spin-triplet, color-singlet states, with
the Dirac and color traces handled by the package
\textsc{FeynCalc}~\cite{Metig:1991}. In the next step, we proceed to
expand the integrand in powers of quark relative momenta, $q$, to
project out the leading $S$-wave and $P$-wave
orbital-angular-momentum contributions.

Making expansion in $q$ prior to carrying out the loop integration,
in the spirit of {\it method of region}~\cite{Beneke:1997zp},
amounts to directly deducing the NRQCD short-distance coefficients,
{\it i.e.}, the contributions solely stemming from the {\it hard}
region ($k^2 \ge m_c^2$). In practice, this procedure is far simpler
than the conventional perturbative matching procedure, since by this
way we will no longer be distracted by the effects from the
low-energy regions, {\it e.g.}, from the {\it soft} ($k^\mu\sim
mv$), or {\it potential} ($k^0\sim mv^2, |{\bf k}|\sim mv$) regions.
Consequently, with this strategy, one will never confront Coulomb
singularity in the one-loop integrals.

In Ref.~\cite{Bodwin:2008nf}, an all-order-in-$\alpha_s$ proof for
exclusive quarkonium production has been outlined in the NRQCD
factorization context. It is argued that, at lowest order in $v$ and
to all orders in $\alpha_s$, NRQCD factorization holds for $S$-wave
plus $P$-wave quarkonia production in $e^+e^-$ annihilation. In
particular, this implies that the NRQCD short-distance coefficients
associated with $e^+e^-\to J/\psi+\chi_{cJ}$ should be free from any
infrared singularity at NLO in $\alpha_s$~\footnote{An uncanceled
infrared singularity has been discovered in NLO calculation for the
$B\to K\chi_{cJ}$~\cite{Song:2002mh,Song:2003yc}, which has been
interpreted as a failure of NRQCD factorization. This symptom hints
that the lower-energy degrees of freedom, the {\it ultrasoft}
($k\sim mv^2$) region, has been erroneously missed, and one should
invoke the even lower energy effective theory, the potential NRQCD,
to remedy this problem. Once the ultrasoft mode from higher Fock
state $|c\bar{c}(^3S_1^{(8)})g\rangle$ is consistently taken into
account, this infrared divergence can be
tamed~\cite{Beneke:2008pi}.}. Indeed, an earlier NLO perturbative
calculation for $e^+e^-\to J/\psi+\chi_{c0}$ has explicitly verified
this pattern~\cite{Zhang:2008gp}.

At this stage, we apply the specifically-designed helicity
projection operators to project out 10 corresponding helicity
amplitudes. This operation brings forth great simplification, for
all the polarization vectors (tensors) have been eliminated, and the
numerators in loop integrals now become the Lorentz scalars
comprised entirely of external and loop momenta. Note these
integrals in general contain propagators with quadratic power due to
the projection of the $P$-wave state.

To proceed, we use the \textsc{Mathematica} package
\textsc{FIRE}~\cite{Smirnov:2008iw} and our self-written
\textsc{Mathematica} code to reduce the general higher-point
one-loop integrals into a set of Master Integrals (MI). Fortunately,
as a great bonus of having Taylor-expanded the integrand in powers
of $q$, together with having applied the helicity projectors, there
are only three linearly-independent Lorentz scalars in loop
integrals: $l^2$, $l\cdot P_1$ and $l\cdot P_2$, where $l$, $P_1$,
$P_2$ stand for the loop momentum, the momenta carried by the
$c\bar{c}(^3S_1^{(1)})$ pair and by the $c\bar{c}(^3P_J^{(1)})$
pair, respectively. Thanks to the integration-by-part algorithm
built in \textsc{FIRE}, it turns out that all the involved MI become
just the usual 2-point and 3-point scalar integrals. All the
encountered 3-point scalar integrals can be found in Appendix of
Ref.~\cite{Gong:2007db}. We have used the package
\textsc{LoopTools}~\cite{Hahn:1998yk} to numerically check the
correctness of these integrals.

Throughout this calculation we adopt dimensional regularization to
regularize both the ultraviolet (UV) and infrared (IR)
singularities. In accordance with the LSZ reduction formula, one
needs to multiply the tree-level amplitude by $(\sqrt{Z_c})^4$,
where $Z_c$ denotes the residue of the charm quark propagator near
its pole. This contribution is represented by
Fig.~\ref{feynman:diagrams}$f)$. In Feynman gauge, the residue is
given by
\beq
Z_c = 1 - C_F \frac{\alpha_s}{4\pi} \left[ \frac{1}{\epsilon_{\rm
UV}} + \frac{2}{\epsilon_{\rm IR}} - 3\gamma_E +
3\ln\frac{4\pi\mu^2}{m_c^2} + 4\right]+ \mathcal{O}(\alpha_s^2),
\label{Z-c}
\eeq
where $\gamma_E$ is the Euler constant, and $C_F= {N_c^2-1\over 2
N_c}$.

In addition, we also need to replace the bare charm quark mass and
the bare QCD coupling constant in the tree-level amplitude by
\begin{subequations}
\bqa
m_c^{\rm bare} &=& m_c \bigg[ 1 - 3 C_F {\alpha_s \over 4\pi} \left(
\frac{1}{\epsilon_{\rm UV}} - \gamma_E + \ln {4\pi\mu^2 \over m_c^2}
+ \frac{4}{3}\right) + \mathcal{O}(\alpha_s^2) \bigg],
\\
\alpha_s^{\rm bare} &=& \alpha_{s}(\mu)^{\overline{\rm MS}} \bigg[1
- \beta_0 \frac{\alpha_s}{4\pi} \left( \frac{1}{\epsilon_{\rm UV}} -
\gamma_E + \ln(4\pi)\right) + \mathcal{O}(\alpha_s^2)\bigg].
\eqa
\end{subequations}
where $\beta_0={11\over 3}N_c-{2\over 3}n_f$ is the one-loop
coefficient of the QCD $\beta$ function, and $n_f=4$ denotes the
number of active quark flavors.

When adding the contributions from all the diagrams, we find that
the ultimate NLO expressions for each of the 10 helicity amplitudes
are UV and IR finite.

\subsection{Analytic expressions of NLO helicity amplitudes}

For clarity, we parameterize the NLO helicity amplitude as follows:
\bqa
{{\mathcal A}^J}^{(1)}_{\lambda_1,\lambda_2}  &=&  {\alpha_s \over
\pi} \, K^J_{\lambda_1,\lambda_2}\left(r,{\mu^2\over s}\right)\,
{{\mathcal A}^J}^{(0)}_{\lambda_1,\lambda_2},
\label{NLO:helic:ampl:parametrization}
\eqa
where the LO helicity amplitude ${\mathcal A}^{(0)}$ is defined in
(\ref{LO:helic:ampl:parametrization}). The dimensionless quantity
$K^J_{\lambda_1,\lambda_2}$, a function of $r$ and the
renormalization scale $\mu$, can be regarded as the reduced NLO
helicity amplitude. It necessarily encompasses all the loop
dynamics. Since helicity selection rule has been tacitly embodied in
${\mathcal A}^{(0)}$ in (\ref{NLO:helic:ampl:parametrization}), we
expect that the $K$ functions will exhibit slower asymptotic
decrease than any power-law scaling in $r$.

\begin{figure}[tbH]
\begin{center}
\includegraphics[scale=0.63]{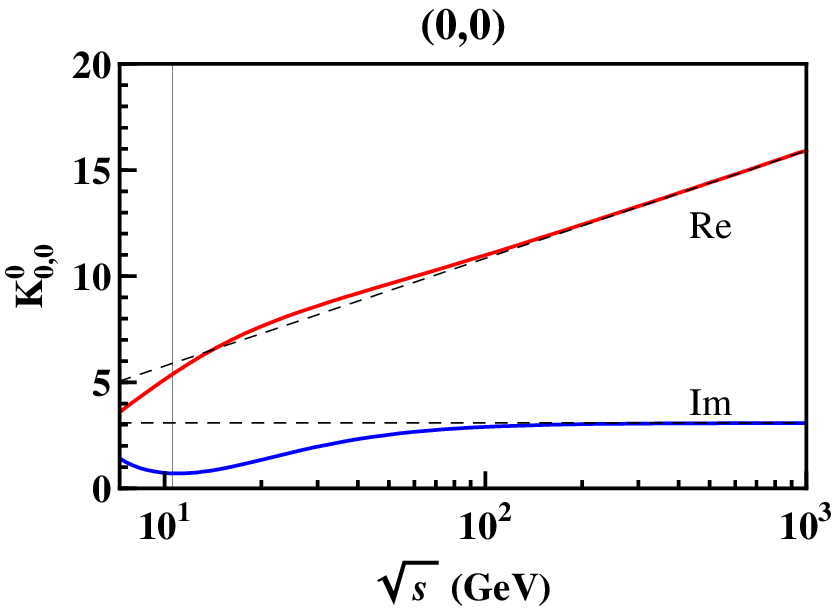}
\includegraphics[scale=0.63]{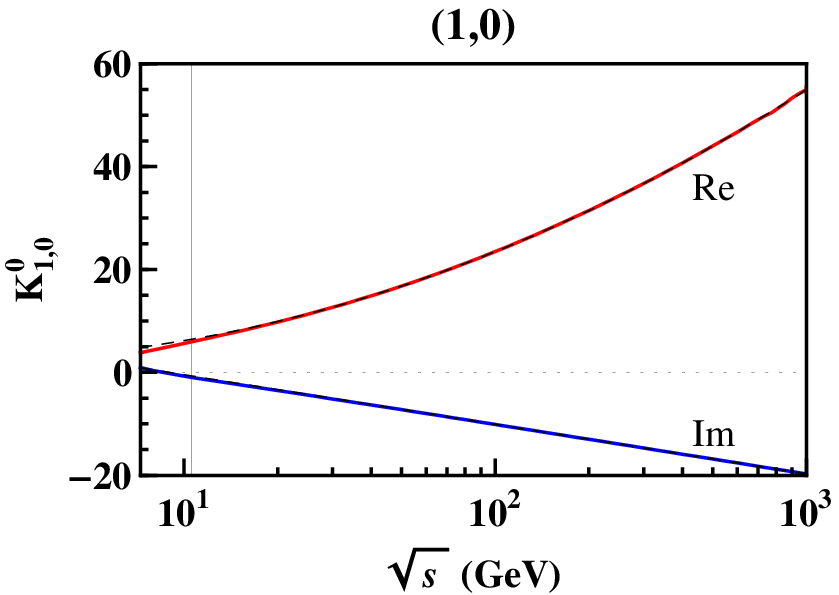}
\caption{The profiles of two reduced NLO helicity amplitudes
$K^0_{\lambda_1,\lambda_2}$ (for $\gamma^*\to J/\psi+\chi_{c0}$)
with $\sqrt{s}$ varied in a wide range. We take $\mu={\sqrt{s}\over
2}$ and $m_c=$ 1.5 GeV. The solid curves correspond to the exact
results, and the dashed curves represent the asymptotic ones as
given in (\ref{K:asymp:NLO:Jpsi:chic0}). The vertical mark is placed
at the $B$ factory energy $\sqrt{s} = 10.58$ GeV.
\label{K:function:jpsi:chic0} }
\end{center}
\end{figure}

\begin{figure}[tbH]
\begin{center}
\includegraphics[scale=0.63]{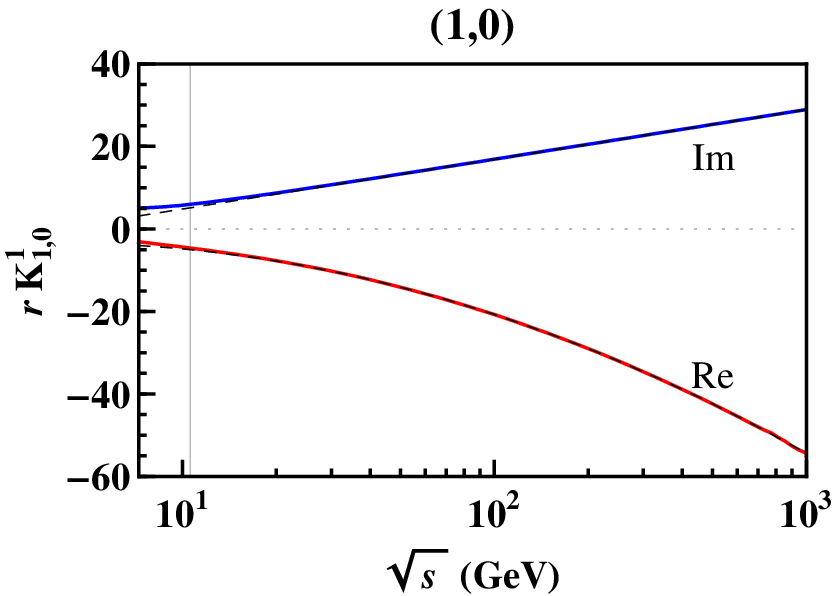}
\includegraphics[scale=0.63]{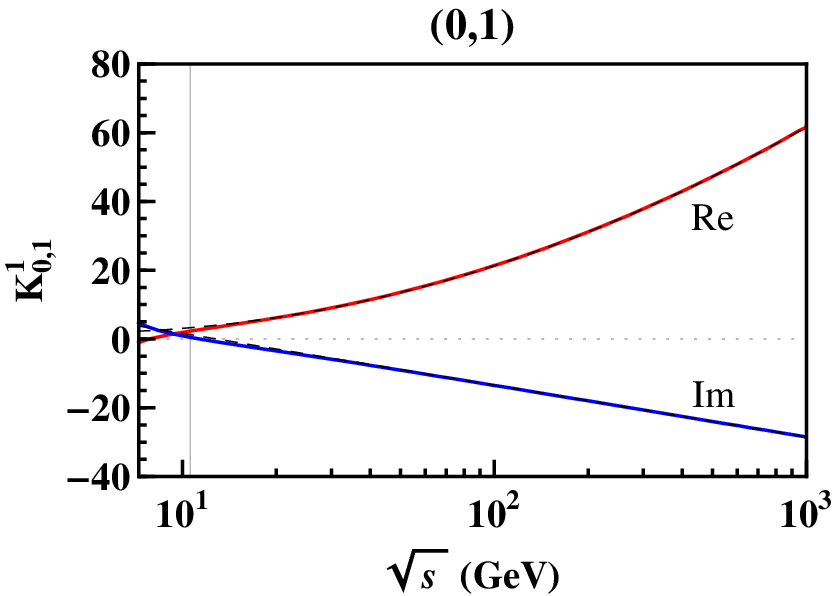}
\includegraphics[scale=0.63]{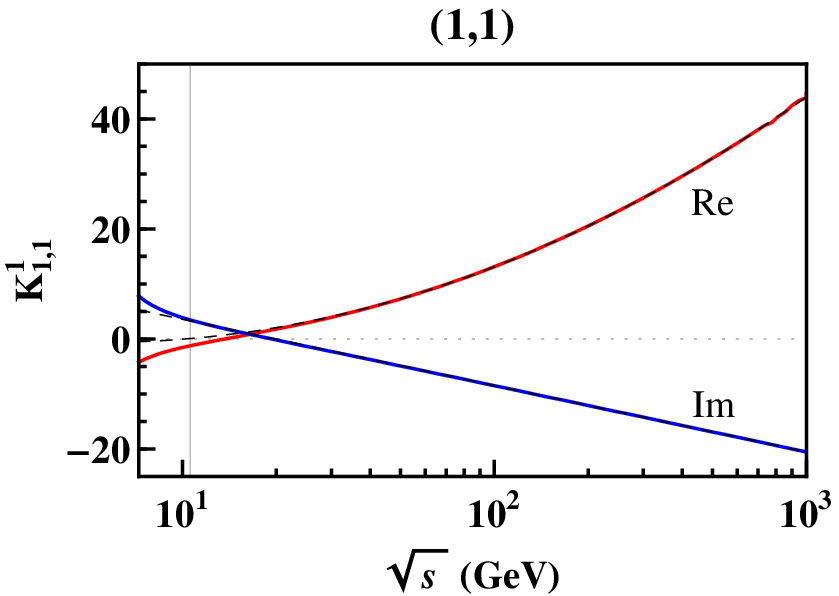}
\caption{The profiles of three reduced NLO helicity amplitudes
$K^1_{\lambda_1,\lambda_2}$ (for $\gamma^*\to J/\psi+\chi_{c1}$).
The asymptotic curves are taken from (\ref{K:asymp:NLO:Jpsi:chic1}).
The parameters are the same as in Fig.~\ref{K:function:jpsi:chic0}.
\label{K:function:jpsi:chic1} }
\end{center}
\end{figure}

\begin{figure}[tbH]
\begin{center}
\includegraphics[scale=0.63]{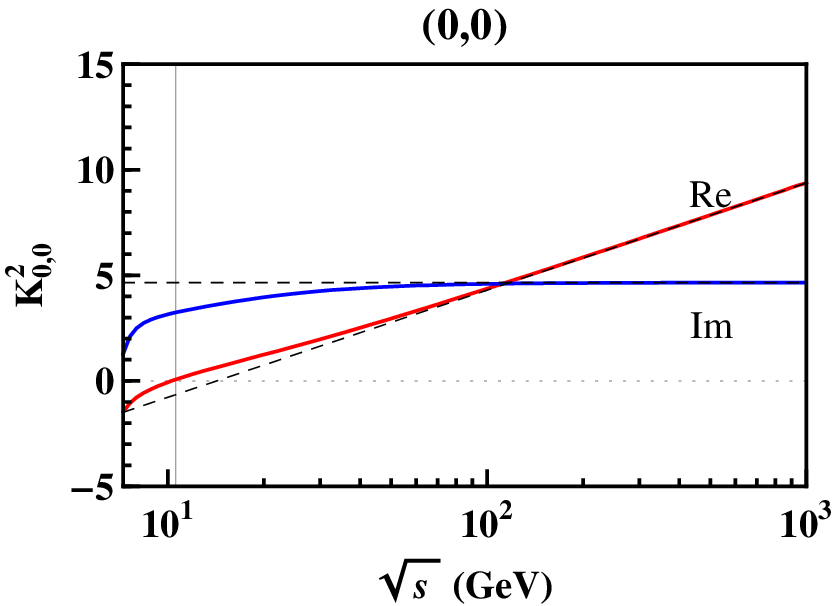}
\includegraphics[scale=0.63]{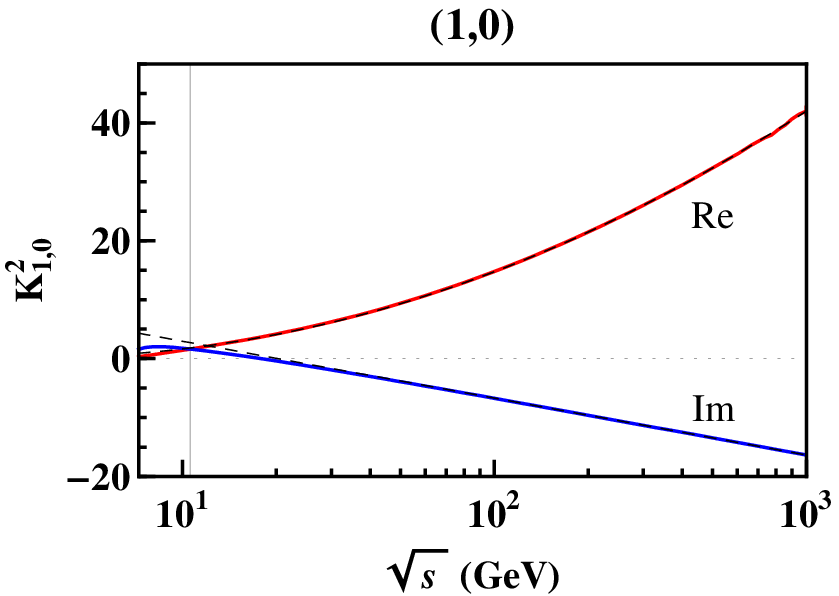}
\includegraphics[scale=0.63]{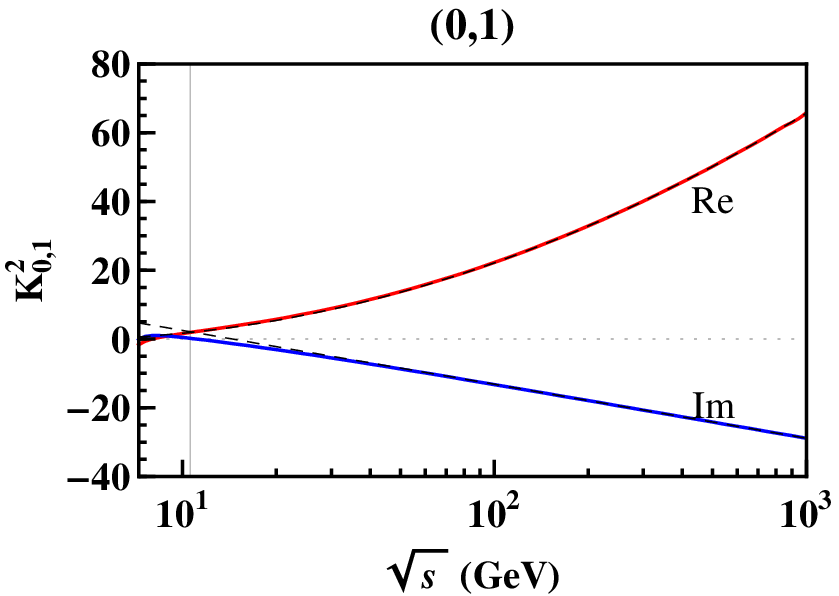}
\includegraphics[scale=0.63]{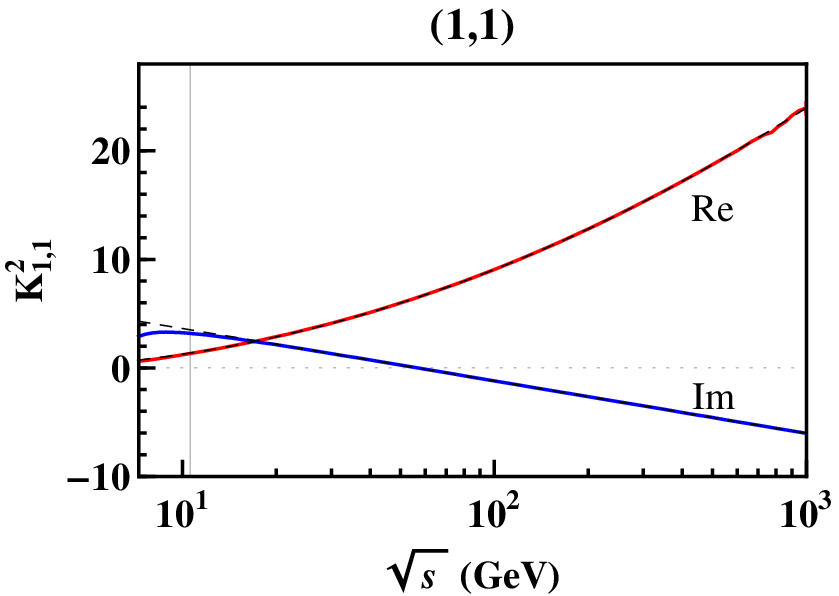}
\includegraphics[scale=0.63]{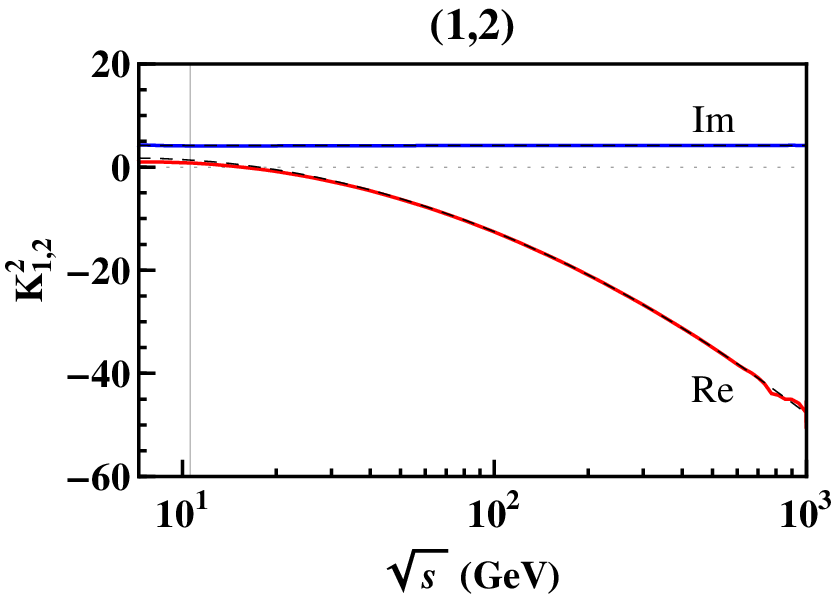}
\caption{The profiles of five reduced NLO helicity amplitudes
$K^2_{\lambda_1,\lambda_2}$ (for $\gamma^*\to J/\psi+\chi_{c2}$).
The asymptotic curves are taken from (\ref{K:asymp:NLO:Jpsi:chic2}).
The parameters are the same as in Fig.~\ref{K:function:jpsi:chic0}.
\label{K:function:jpsi:chic2} }
\end{center}
\end{figure}

The reduced NLO helicity amplitudes $K$ are in general
complex-valued. Their analytic expressions are somewhat lengthy and
will not be reproduced here. On the other hand, in
Figs.~\ref{K:function:jpsi:chic0}, \ref{K:function:jpsi:chic1}, and
\ref{K:function:jpsi:chic2}, we explicitly display their profiles
over a wide range of $r$ for each of the 10 helicity amplitudes.

Combining (\ref{LO:helic:ampl:parametrization}),
(\ref{NLO:helic:ampl:parametrization}) and
(\ref{unpol:cross:section:Jpsi:chicJ}), we can deduce the NLO
perturbative correction to the polarized cross section:
\bqa
& & \sigma^{(1)}[J/\psi(\lambda_1) + \chi_{cJ}(\lambda_2)] =
2\left({\alpha_s\over \pi}\right) {\rm Re}\bigg\{
K^J_{\lambda_1,\lambda_2}\left(r,{\mu^2\over s}\right)
\bigg\}\,\sigma^{(0)}[J/\psi(\lambda_1) + \chi_{cJ}(\lambda_2)],
\label{pol:cross:section:NLO:alphas}
\eqa
where $\sigma^{(0)}$ is given in
(\ref{pol:cross:section:LO:alphas}). To the desired NLO accuracy,
the imaginary part of the $K$ function does not contribute. We now
have improved prediction $\sigma_{\rm NLO} =
\sigma^{(0)}+\sigma^{(1)}$.

It is theoretically curious to know the asymptotic behaviors of
these reduced helicity amplitudes in the limit $\sqrt{s}\gg m_c$. As
was mentioned before, we anticipate to see the logarithmic scaling
violation at NLO to the naive power-law scaling rule
(\ref{helicity:selection:rule}). Moreover, performing the asymptotic
expansion in NRQCD short-distance coefficients is theoretically very
appealing, since it is equivalent to disentangling the effects
occurring at the ``hard" scale (virtuality $\sim s$) from the
lower-energy ``collinear/soft" effects (virtuality $\sim m_c^2$), by
which one can intimately link the NRQCD factorization approach and
the light-cone approach~\cite{Jia:2008ep,Jia:2010fw}. Such
asymptotic expansion has been carried out for several
double-quarkonium production processes at one-loop
level~\cite{Jia:2007hy,Gong:2008ue,Jia:2010fw} and some general
patterns have been recognized.

The asymptotic expressions of two reduced NLO helicity amplitudes
for $J/\psi+\chi_{c0}$ are
\begin{subequations}
\bqa
& & K^0_{0,0} \left( r,{\mu^2 \over s} \right)_{\rm asym} = -{1\over
3} (4-\ln2) \ln r + {\beta_0 \over 4} \bigg( \ln {4 \mu^2\over s} +
{8\over 3}\bigg)
\label{asymp:NLO:K000}
\nn \\
&& - \frac{1}{18} ( 46 + \pi^2 - 40\ln2 + 33 \ln^2 2) + {i\pi\over
4} \bigg(\beta_0 -{16\over 3}+ {4\over 3}\ln2 \bigg),
\\
& &  K^0_{1,0}\left( r,{\mu^2 \over s} \right)_{\rm asym} = {1\over
3} \ln^2 r - {1 \over 108}(139-104\ln2)\ln r + {\beta_0 \over 4}
\bigg( \ln{4 \mu^2 \over s} + {17\over 9} \bigg)
\nn \\
&& - {1\over 54}\bigg( 161 + {8\pi^2\over 3}- {495\over 2}\ln2 +
100 \ln^2 2 \bigg) + {i\pi \over 4} \bigg( {8\over 3}\ln r +
\beta_0 - {1\over 27}(139 -104\ln 2)\bigg).
\nn\\
\eqa
\label{K:asymp:NLO:Jpsi:chic0}
\end{subequations}

The asymptotic expressions of the three reduced NLO helicity
amplitudes for $J/\psi+\chi_{c1}$ read
\begin{subequations}
\bqa
& & r\,K^1_{1,0}\left( r,{\mu^2 \over s} \right)_{\rm asym} =
-{1\over 12} \Bigg\{ 5 \ln^2r + (7-2\ln2)\ln r -19 + 2\pi^2 + 75\ln
2- 21 \ln^2 2
\label{asymp:NLO:K110}
\nn \\
&& + i\pi ( 10\ln r + 7 - 2\ln2 )\Bigg\},
\\
& & K^1_{0,1}\left( r,{\mu^2 \over s} \right)_{\rm asym} = {1\over
24}\Bigg\{ \frac{25}{2} \ln^2 r - (46-99\ln2)\ln r + 6\beta_0 \bigg(
\ln{4\mu^2\over s}  + {13\over 6}\bigg)
\nn \\
&& - {1\over 6} (616 + 74\pi^2 - 1696\ln2 + 303 \ln^22 ) +
i\pi(25\ln r + 6 \beta_0 -46 + 99\ln2 )\Bigg\},
\\
& & K^1_{1,1} \left( r,{\mu^2 \over s} \right)_{\rm asym} = {1\over
24}\Bigg\{ 10 \ln^2r + 2(1+17\ln2)\ln r + 6\beta_0 \bigg(
\ln{4\mu^2\over s}  + {13\over 6}\bigg)
\nn \\
&& - \frac{1}{3}( 266 + 7\pi^2 - 128\ln 2 + 147 \ln^2 2) + 2 i\pi(
10\ln r + 3\beta_0 + 1+ 17\ln2) \Bigg\}.
\eqa
\label{K:asymp:NLO:Jpsi:chic1}
\end{subequations}

The asymptotic expressions of the five reduced NLO helicity
amplitudes for $J/\psi+\chi_{c2}$ are
\begin{subequations}
\bqa
& & K^2_{0,0}\left( r,{\mu^2 \over s} \right)_{\rm asym} = -{1\over
3}(4-\ln2) \ln r + {\beta_0 \over 4} \bigg(\ln{4\mu^2 \over s} +
{8\over 3} \bigg)
\nn \\
&& -\frac{1}{18} ( 64 + \pi^2 + 104\ln 2 + 33 \ln^22) +
 \frac{i\pi}{4} \bigg( \beta_0 - {10 \over 3} + {4\over 3}\ln2 \bigg),
\label{asymp:NLO:K200}
\\
& & K^2_{0,1} \left( r,{\mu^2 \over s} \right)_{\rm asym} = {1\over
12} \Bigg\{ \frac{13}{2}\ln^2 r - (22-43\ln2) \ln r + 3 \beta_0
 \bigg(\ln{4\mu^2 \over s} +
{8\over 3} \bigg)
\nn \\
&& - \frac{1}{6} ( 284 + 30\pi^2 - 380\ln2 + 159 \ln^2 2) + i\pi
(13\ln r + 3 \beta_0 - 14 + 43\ln2 )\Bigg\},
\\
& & K^2_{1,0}\left( r,{\mu^2 \over s} \right)_{\rm asym} = {1\over
6} \Bigg\{2\ln^2 r +\frac{1}{6}(5+8\ln2) \ln r + {3\over 2}\beta_0
\bigg(
 \ln {4\mu^2\over s} + {7\over 3} \bigg)
\nn\\
&& -\frac{1}{18} ( 291-8\pi^2 + 171 \ln 2 + 312\ln^2 2 ) + i\pi
\bigg( 4\ln r + {3\over 2} \beta_0 + {11\over 6} + {4 \over 3} \ln2
\bigg)\Bigg\},
\\
& & K^2_{1,1} \left( r,{\mu^2 \over s} \right)_{\rm asym}  = {1\over
24} \Bigg\{ 4\ln^2 r - (46-62\ln2) \ln r + 6 \beta_0 \bigg(
\ln{4\mu^2\over s} + {13\over 6} \bigg )
\nn\\
&& - \frac{1}{3}(274 + 27\pi^2 - 316\ln 2 + 9\ln^2 2) + i\pi (8\ln r
+ 6 \beta_0 -46 +62\ln 2)\Bigg\},
\\
& & K^2_{1,2} \left( r,{\mu^2 \over s} \right)_{\rm asym} = -{1\over
4} \Bigg\{ 2\ln^2 r + \frac{2}{3}(1+13\ln2) \ln r - \beta_0 \bigg(
\ln{4\mu^2\over s} + {5\over 3} \bigg)
\nn\\
&& + \frac{1}{9} ( -7\pi^2 + 140 - 104\ln 2 + 237\ln^2 2)- i\pi
\bigg(\beta_0 + 3 - {26\over 3} \ln2 \bigg)\Bigg\}.
\eqa
\label{K:asymp:NLO:Jpsi:chic2}
\end{subequations}

Now we can make several interesting observations from
Eqs.~(\ref{K:asymp:NLO:Jpsi:chic0}), (\ref{K:asymp:NLO:Jpsi:chic1}),
(\ref{K:asymp:NLO:Jpsi:chic2}). One confirms that the scaling
violation is indeed of the logarithmic form. For the
hadron-helicity-conserving channels such as
$J/\psi(0)+{\chi_{c0,2}}(0)$, the leading scaling behavior of the
real part of the $K$ function is governed by a single logarithm of
$r$; for all the remaining helicity-suppressed channels, the leading
asymptotic behaviors of the respective $K$ functions are all
proportional to $\ln^2 r$.

This pattern lends further support for the speculation made in
Ref.~\cite{Jia:2010fw}: The hard exclusive processes involving
double-quarkonium at leading twist can only accommodate the single
collinear logarithm at one-loop, while those beginning with the
higher twist contributions are always plagued with double
logarithms. For example, the NLO NRQCD short-distance coefficients
of the helicity-suppressed reactions $e^+e^-\to J/\psi+\eta_c$ and
$\eta_b\to J/\psi J/\psi$ are both found to contain the
double-logarithm term~\cite{Jia:2010fw,Gong:2008ue}.

Most of the studied double-charmonium production processes are of
the helicity-suppressed type. In this regard, the helicity channels
$e^+e^-\to J/\psi(0)+{\chi_{c0,2}}(0)$ constitute the rather rare
examples that the leading-twist contribution dominates. In such a
situation, by resorting to the leading-twist collinear factorization
theorem, one can employ the light-cone approach to efficiently
reproduce the asymptotic expressions given in (\ref{asymp:NLO:K000})
and (\ref{asymp:NLO:K200}), very much like what is achieved for
$B_c$ electromagnetic form factor at NLO in
$\alpha_s$~\cite{Jia:2010fw}. Note that the single logarithm in
these two channels have identical coefficient $\propto 4-\ln 2$.
This coefficient can be readily reconstructed in light-cone
approach, with the aid of the Efremov-Radyushkin-Brodsky-Lepage
(ERBL) evolution equation~\cite{Lepage:1979zb,Efremov:1979qk}.
Moreover, following the strategy of \cite{Jia:2008ep}, by employing
this evolution equation, one can systematically identify and resum
the leading collinear logarithms in these amplitudes to all orders
in $\alpha_s$.

In contrast, it remains to be an open challenge for light-cone
approach to reproduce these process-dependent double logarithms
appearing in NLO NRQCD short-distance coefficients, which seem to
have resulted from the overlap between collinear and endpoint
singularities~\cite{Jia:2010fw}. Note that the end-point singularity
is a long-standing problem in light-cone framework, which has
essentially hindered our capability of performing the complete NLO
perturbative calculation beyond leading-twist using this approach.
Perhaps the first major progress is to successfully reproduce the
asymptotic expressions related with those $(0,1)$ and $(1,0)$
helicity channels, where only twist-3 effects need be considered.

For reader's convenience, all the asymptotic results of the reduced
helicity amplitudes are also shown in
Figs.~\ref{K:function:jpsi:chic0}, \ref{K:function:jpsi:chic1},
\ref{K:function:jpsi:chic2}, in juxtapose with the corresponding
exact NLO results. As can be seen clearly, for those higher-twist
helicity channels, the asymptotic results tend to converge with the
exact ones decently well at relatively lower $\sqrt{s}$ (say, at $B$
factory energy) than for the leading-twist channels. This seems to
be a general feature, which has also been observed in several other
double charmonium production
processes~\cite{Jia:2007hy,Gong:2008ue,Jia:2010fw}.

To conclude this section, we mention a peculiar phenomenon
affiliated with the helicity channel $\gamma^*\to
J/\psi(1)+\chi_{c1}(0)$. Recall that this helicity amplitude at LO
has been suppressed by an unwanted factor of $r$, as can be seen in
(\ref{c:function:LO:chic1}). Nevertheless, as one may readily tell
from (\ref{asymp:NLO:K110}), at NLO in $\alpha_s$, this helicity
amplitude recovers the correct power-law scaling as dictated by the
helicity selection rule! This implies that at very high energy, the
NLO contribution is far more significant than the LO piece for this
helicity channel, despite its extra suppression by $\alpha_s$. It
might be also worth noting that, unlike all the other asymptotic
expressions, the renormalization logarithm $\beta_0\ln\mu^2/s$ is
absent in (\ref{asymp:NLO:K110}). This is very similar to what is
found in NLO perturbative correction to $\eta_b\to J/\psi
J/\psi$~\cite{Gong:2008ue}.

\section{Phenomenology}
\label{phenomenology}
\setcounter{equation}{0}

With our NLO calculations completed, following the formulas
(\ref{pol:cross:section:LO:alphas}) and
(\ref{pol:cross:section:NLO:alphas}), we are ready to carry out a
detailed analysis for the processes $e^+e^-\to J/\psi(\psi') +
\chi_{cJ}$ and confront the $B$ factory measurements. In our
numerical analysis, we set $\sqrt{s}=10.58$ GeV, $m_c=1.5$ GeV. The
electromagnetic fine structure constant is chosen as
$\alpha(\sqrt{s}) = 1/130.9$~\cite{Bodwin:2007ga}. The running QCD
strong coupling constant is evaluated by using the two-loop formula
with $\Lambda^{(4)}_{\overline{\rm MS}}= 0.338$
GeV~\cite{Zhang:2005cha,Gong:2007db}. The nonpertubative input
parameters, {\it i.e.}, the wave function at the origin for $J/\psi$
($\psi^\prime$), as well as the first derivative of the radial wave
function at the origin for $\chi_{cJ}$, bear a fair amount of
uncertainties. Their values have been compiled in
Ref.~\cite{Eichten:1995}, which are estimated from several different
potential models. We choose to use those given by the
Buchm\"{u}ller-Tye potential model~\cite{Buchmuller:1981}: $\vert
R_{J/\psi}(0) \vert^2= 0.81 \; {\rm GeV}^3$,  $\vert
R_{\psi'}(0)\vert^2 = 0.529 \;{\rm GeV}^3$, and $\vert
R'_{\chi_{cJ}}(0)\vert^2 = 0.075 \;{\rm GeV}^5$.

Another important source of uncertainty for the NLO predictions
comes from the scale setting for the strong coupling constant. As is
well known, this scale ambiguity is a typical nuisance of NRQCD
factorization approach, reflecting the fact that two disparate hard
scales, $\sqrt{s}$ and $m_c$, are entangled together in NRQCD
short-distance coefficients. In contrast, the light-cone approach,
when armed with the idea of refactorization, can efficiently
disentangle these two scales, so one naturally expects that the
scale ambiguity will be greatly reduced.

While it may sound natural to choose the renormalization scale $\mu$
to be around the highest scale $\sqrt{s}$, however, as is clearly
illustrated in \cite{Jia:2010fw}, it is rather unappealing to set
the scales entering all $\alpha_s$ in NLO short-distance coefficient
to be around $\sqrt{s}$ unanimously. The reason is that a part of
NLO correction comes from the loop region with lower virtuality, and
as such can be identified with the NLO perturbative correction to
the charmonium decay constant, so the corresponding $\alpha_s$
should definitely be affiliated with a scale around $2 m_c$ rather
than $\sqrt{s}$.

Since it is not possible to completely solve the scale ambiguity
problem within the confines of the NRQCD factorization approach, we
proceed to estimate the cross section by assigning all the occurring
$\alpha_s$ with a common scale, $\mu$, and choosing $\mu=\sqrt{s}/2$
and $\mu=2 m_c$, respectively. It is hoped that the less biased
results interpolate between these two sets of predictions.

\begin{figure}[tbh]
\begin{center}
\includegraphics[scale=0.63]{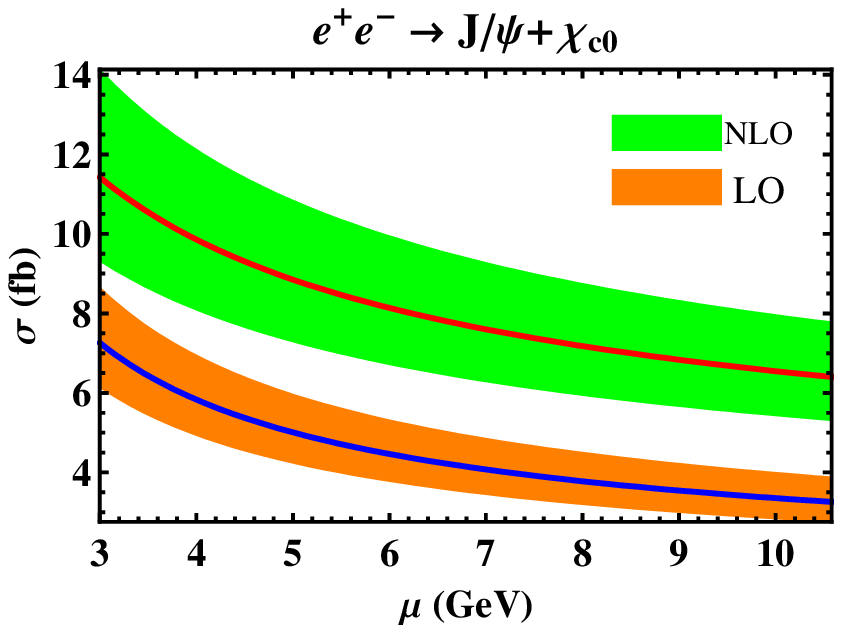}
\includegraphics[scale=0.63]{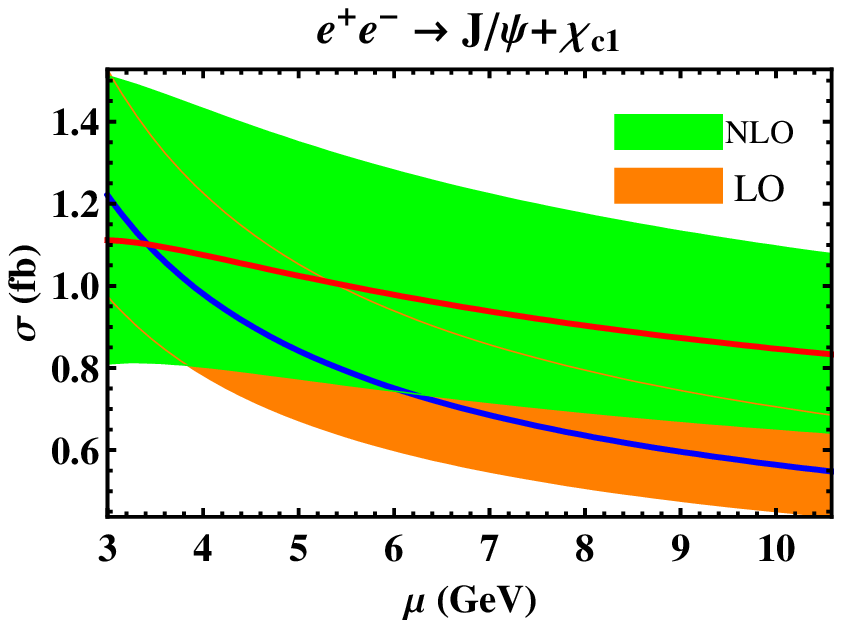}
\includegraphics[scale=0.63]{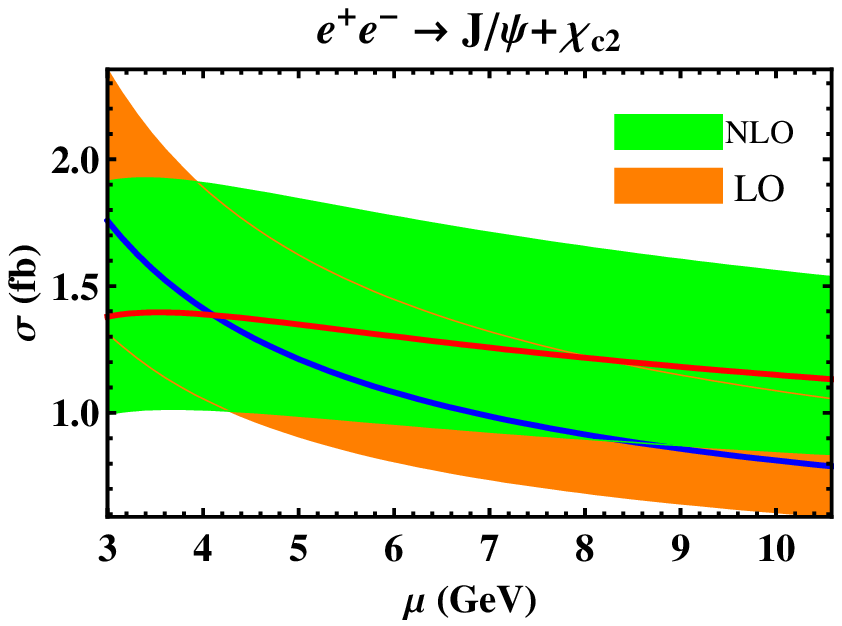}
\caption{The $\mu$-dependence of LO and NLO cross sections for
$e^+e^-\to J/\psi+\chi_{cJ}$ ($J=0,1,2$) at $\sqrt{s} = 10.58$ GeV.
The uncertainty band is due to sliding $m_c$ between $1.4$ GeV and
$1.6$ GeV, where the central curves correspond to $m_c = 1.5 $ GeV.
\label{Plot:mu:variation:mc:error:band} }
\end{center}
\end{figure}

\begin{figure}[tbh]
\begin{center}
\includegraphics[scale=0.63]{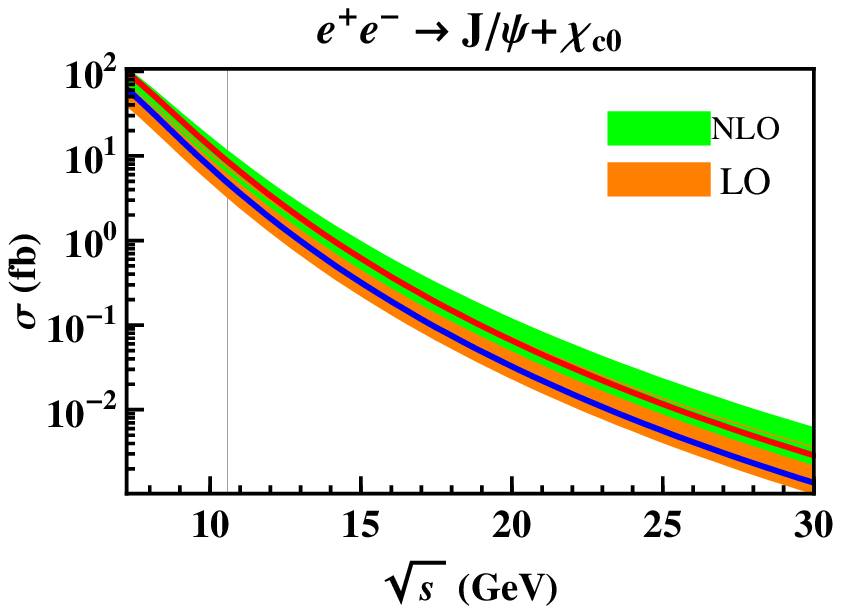}
\includegraphics[scale=0.63]{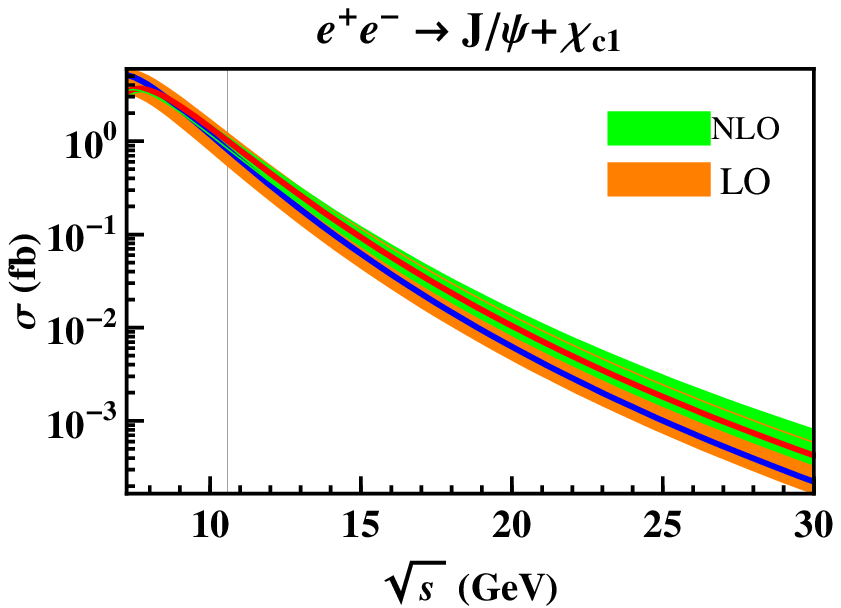}
\includegraphics[scale=0.63]{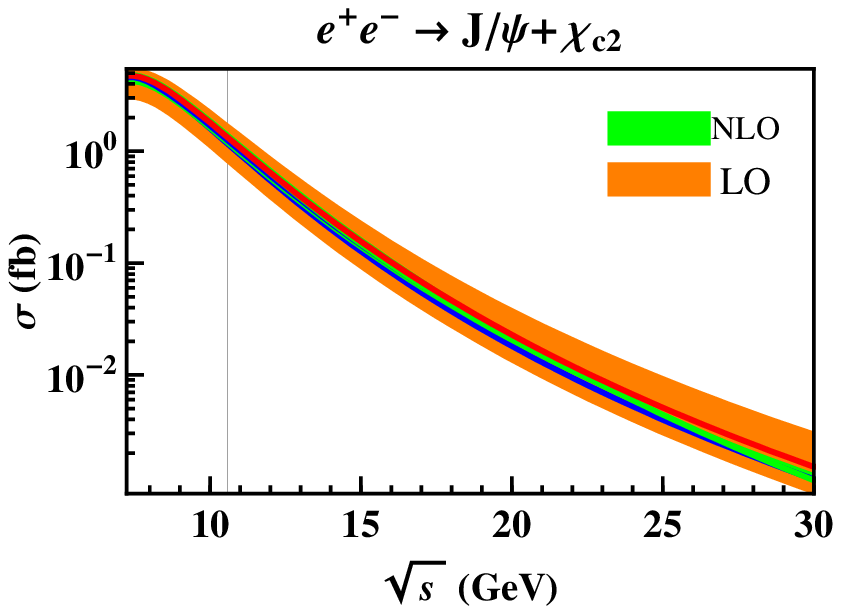}
\caption{The LO and NLO cross sections for $e^+ e^-\to
J/\psi+\chi_{cJ}$ ($J=0,1,2$) as a function of $\sqrt{s}$. The
uncertainty band is obtained by varying $\mu$ from $2m_c$ to
$\sqrt{s}$, where the central curves represent the default choice
$\mu = {\sqrt{s}\over 2}$.
\label{Plot:sqrst:variation:mu:error:band}}
\end{center}
\end{figure}

To develop a concrete feel about the scale dependence, in
Fig.~\ref{Plot:mu:variation:mc:error:band} we explicitly show the
$\mu$-dependence of the LO and NLO total cross sections for the
processes $e^+e^-\to J/\psi+\chi_{cJ}$ ($J=0,1,2$) at $B$ factory
energy. As can be clearly seen, including the NLO correction has
notably reduced the scale dependence for $J/\psi+\chi_{c1,2}$,
whereas of little impact for $J/\psi+\chi_{c0}$. Also in
Fig.~\ref{Plot:mu:variation:mc:error:band} we have examined the
$m_c$-dependence of the cross section, which is reflected in the
error band by varying $m_c$ from $1.4$ GeV and $1.6$ GeV.

In Fig.~\ref{Plot:sqrst:variation:mu:error:band}, we also plot the
LO and NLO total cross sections for $e^+ e^-\to J/\psi+\chi_{cJ}$
($J=0,1,2$) as a function of $\sqrt{s}$. The error band is obtained
by sliding $\mu$ from $2m_c$ to $\sqrt{s}$, where the central curves
represent the default choice $\mu = {\sqrt{s}\over 2}$. For the LO
predictions, all the $J/\psi+\chi_{cJ}$ channels have comparable
widths of the error bands. But the NLO results for
$J/\psi+\chi_{c2}$ exhibit a significantly narrower band compared to
those for $J/\psi+\chi_{c0,1}$.

\begin{table}[tbH]
\caption{Polarized cross sections for each helicity channel and
unpolarized (total) cross sections (in unit of fb) for $e^+e^-\to
J/\psi+\chi_{cJ}$ ($J=0, 1, 2$). In the rightmost column, we also
list the K factor for the unpolarized cross sections. We choose the
following input parameters: $m_c = 1.5$ GeV, $\mu=\sqrt{s}/2$,
accordingly, $\alpha_s(\mu)= 0.211$.}
\begin{center}
\begin{tabular}{ccc ccc ccc ccc cccc ccc ccc ccc ccc}
\hlinew{1pt}
    &&&  \ \    &&&\ \ \ \  $\sigma_{(0,0)}$\ \ \  &&& \ \ \
    $\sigma_{(1,0)}$ \ \ \  &&& \ \ \ $\sigma_{(0,1)}$  \ \ \
    &&& \  \ \  $\sigma_{(1,1)}$\ \ \   &&&\ \ \  $\sigma_{(1,2)}$ \ \ \
     &&& \ \ \  $\sigma_{\rm tot}$\ \ \ &&&\ \ \  K \ \ \ \ \\
\hline

\ \ \multirow{2}{*}{$J/\psi+\chi_{c0}$}\ \
 &&& LO        &&& $1.11$  &&& $1.86$&&&    --    &&&    --
 &&&     --  &&& $4.83$ &&&\multirow{2}{*}{$1.79$}\\
 &&& NLO      &&&  $1.92$  &&& $3.35$&&&    --    &&&    --   &&&     --  &&&  $8.62$ \\
\hline

\multirow{2}{*}{$J/\psi+\chi_{c1}$}
&&& LO         &&&   --   &&&$0.0012$&&&$0.37$&&&$0.033$&&&     --
&&&$0.81$ &&&\multirow{2}{*}{$1.24$}\\
 &&& NLO       &&&   --   &&&$-0.0078$ &&&$0.49$ &&&$0.028$
 &&&     --  &&&$1.01$\\
\hline

\multirow{2}{*}{$J/\psi+\chi_{c2}$}
&&& LO  &&&$0.43$&&&$0.27$&&&$0.064$&&&$0.033$&&&$0.0023$&&&$1.17$ &&&\multirow{2}{*}{$1.14$}\\
&&& NLO &&&$0.44$&&&$0.33$ &&&$0.081$&&&$0.039$&&&$0.0026$
  &&&$1.33$  \\
\hlinew{1pt}
\end{tabular}
\end{center}
\label{Table-1}
\end{table}

\begin{table}[tbH]
\caption{The same as Table~\ref{Table-1}, except we choose the
different renormalization scale: $\mu=2m_c$, accordingly
$\alpha_s(\mu) = 0.259$.}
\begin{center}
\begin{tabular}{ccc ccc ccc ccc cccc ccc ccc ccc ccc}
\hlinew{1pt}
    &&&  \ \    &&&\ \ \ \  $\sigma_{(0,0)}$\ \ \  &&& \ \ \ $\sigma_{(1,0)}$ \ \ \
    &&& \ \ \ $\sigma_{(0,1)}$  \ \ \
    &&& \  \ \  $\sigma_{(1,1)}$\ \ \   &&&\ \ \  $\sigma_{(1,2)}$ \ \ \
    &&& \ \ \  $\sigma_{\rm tot}$\ \ \ &&&\ \ \  K \ \ \ \ \\
\hline

\ \ \multirow{2}{*}{$J/\psi+\chi_{c0}$}\ \
 &&& LO        &&& $1.67$  &&& $2.80$&&&    --    &&&    --   &&&     --  &&& $7.26$
 &&&\multirow{2}{*}{$1.57$}\\
 &&& NLO      &&&  $2.50$  &&& $4.46$&&&    --    &&&    --   &&&     --  &&&  $11.43$ \\
\hline

\multirow{2}{*}{$J/\psi+\chi_{c1}$}
&&& LO         &&&   --   &&&$0.0017$&&&$0.56$&&&$0.050$&&&     --
&&&$1.22$ &&&\multirow{2}{*}{$0.91$}\\
 &&& NLO       &&&   --   &&&$-0.016$ &&&$0.55$ &&&$0.020$ &&&  --
 &&& $1.11$\\
\hline

\multirow{2}{*}{$J/\psi+\chi_{c2}$}
&&& LO  &&&$0.65$&&&$0.40$&&&$0.097$&&&$0.050$&&&$0.0035$&&&$1.76$ &&&\multirow{2}{*}{$0.78$}\\
&&& NLO &&&$0.40$&&&$0.35$ &&&$0.089$&&&$0.041$&&&$0.0026$
  &&&$1.38$  \\
\hlinew{1pt}
\end{tabular}
\end{center}
\label{Table-2}
\end{table}

\begin{table}[tbH]
\caption{Comparison between our predicted unpolarized cross sections
with the measurements at B factories for $e^+e^-\to
J/\psi(\psi')+\chi_{cJ}$ ($J=0, 1, 2$). The cross sections are in
units of fb. We fix $m_c = 1.5$ GeV, the error is estimated by
varying $\mu$ from $2m_c$ to $\sqrt{s}$, where the central value
refers to $\mu=\sqrt{s}/2$.\label{cross-m}}
\begin{center}
\begin{tabular}{ccc ccc ccc ccc ccc ccc c}
\hlinew{1pt}
&&&  \ \    &&& \textsc{Belle} &&& \textsc{BaBar}  &&& LO prediction &&& NLO prediction  &&& \\
&&&  \ \    &&& $\sigma\times\mathcal{B}_{>2 (0)}$\cite{Abe:2004ww}
&&& $\sigma\times\mathcal{B}_{>2}$\cite{Aubert:2005tj}  &&&  &&&  &&& \\
\hline &&&  $\sigma(J/\psi+\chi_{c0})$    &&& $6.4\pm1.7\pm1.0$ &&&
$10.3\pm 2.5^{+1.4}_{-1.8}$
&&&  $4.83^{+2.43}_{-1.57}$  &&&  $8.62^{+2.80}_{-2.22}$  &&& \\
&&&  $\sigma(J/\psi+\chi_{c1})$    &&& - &&& -  &&&
$0.81^{+0.41}_{-0.26}$  &&& $1.01^{+0.10}_{-0.18}$ &&& \\
&&&  $\sigma(J/\psi+\chi_{c2})$    &&& - &&& -  &&&
$1.17^{+0.59}_{-0.38}$ &&&
$1.33^{+0.04}_{-0.20}$ &&& \\
&&&  $\sigma(J/\psi+\chi_{c1}) + \sigma(J/\psi+\chi_{c2})$
&&& $ < 5.3$ at 90\% CL &&& -  &&&  $1.98^{+1.00}_{-0.64}$ &&&  $2.35^{+0.14}_{-0.38}$ &&& \\
\hline &&&  $\sigma(\psi'+\chi_{c0})$    &&& $12.5\pm3.8\pm3.1$
&&& -  &&&  $2.79^{+1.41}_{-0.91}$  &&&  $4.98^{+1.62}_{-1.28}$  &&& \\
&&&  $\sigma(\psi' + \chi_{c1})$    &&& - &&& -  &&&
$0.47^{+0.24}_{-0.15}$
&&& $0.58^{+0.06}_{-0.10}$ &&& \\
&&&  $\sigma(\psi'+\chi_{c2})$    &&& - &&& -  &&&
$0.68^{+0.34}_{-0.22}$
&&&  $0.77^{+0.03}_{-0.12}$ &&& \\
&&&  $\sigma(\psi'+\chi_{c1}) + \sigma(\psi'+\chi_{c2})$ &&& $< 8.6$
at 90\% CL &&& -  &&&  $1.14^{+0.58}_{-0.37}$
&&&  $1.36^{+0.08}_{-0.22}$ &&& \\
\hlinew{1pt}
\end{tabular}
\end{center}
\label{Table-3}
\end{table}

Table~\ref{Table-1} and Table~\ref{Table-2} tabulate our predictions
to both of the LO and NLO cross sections for $e^+e^-\to
J/\psi+\chi_{cJ}$ $(J=0,1,2)$, with these two sets of $\mu$ chosen
respectively. In addition to the unpolarized (total) cross sections,
we also include the polarized cross sections from each helicity
channel.

Let us first discuss the unpolarized cross sections. For $e^+e^-\to
J/\psi+\chi_{c0}$, the NLO correction has significantly enhanced the
LO prediction, with a K factor of 1.79 and 1.57 respectively,
corresponding to two different choices of $\mu$~\footnote{We find
disagreement with the previous NLO correction calculation for
$e^+e^-\to J/\psi+\chi_{c0}$~\cite{Zhang:2008gp}. When taking the
same input parameters as theirs, we obtain the K factor of 1.57,
while theirs is 2.8.}. In contrast, the NLO corrections to
$e^+e^-\to J/\psi+\chi_{c1,2}$ have a milder impact, even with the
sign uncertain. Concretely speaking, the first setting of $\mu$
tends to increase the LO results modestly, while the second setting
tends to reduce the LO results to some extent. This behavior may be
clearly seen in Fig.~\ref{Plot:mu:variation:mc:error:band}. There
the LO and NLO prediction bands cross with each other for $e^+e^-\to
J/\psi+\chi_{c1,2}$, which implies that the $K$ factor could be
above or below 1, depending on the chosen scale. A similar
understanding can also be achieved in
Fig.~\ref{Plot:sqrst:variation:mu:error:band}. At $\sqrt{s}=10.58$
GeV, one sees from Fig.~\ref{Plot:sqrst:variation:mu:error:band} the
uncertainty band of the NLO predictions for $e^+e^-\to
J/\psi+\chi_{c1,2}$ has been completely submerged inside the band of
the LO predictions. Therefore, depending on whether choosing the
lower or upper bound for LO cross section, the $K$ factor would be
greater or less than 1.

In Table~\ref{Table-3}, we compare our predicted cross sections with
the measurements at B factories for $e^+e^-\to
J/\psi(\psi')+\chi_{cJ}$ $(J=0,1,2)$. One sees that the NLO
perturbative correction to $J/\psi+\chi_{c0}$ really brings the
NRQCD prediction closer to the data, although there exists some
slight tension between \textsc{BaBar} and \textsc{Belle}
measurements. By far, the $e^+e^-\to J/\psi(\psi')+\chi_{c1,2}$
processes have not yet been observed in any experiments. Our
predictions for the $J/\psi(\psi')+\chi_{c1,2}$ cross sections are
compatible with the upper bounds placed by the \textsc{Belle}
experiment. However, even if the large positive NLO correction is
taken into account, our predicted $\psi'+\chi_{c0}$ cross section is
still significantly below the central value of the \textsc{Belle}
measurement. To clarify this puzzling situation, it seems necessary,
and, urgent, for \textsc{BaBar} to perform an independent
measurement for this process to see whether it confirms or
disconfirms the \textsc{Belle} results.

It seems foreseeable that, in future Super $B$ factory,
experimentalists may be able to measure some of the polarized cross
sections for the $e^+e^-\to J/\psi+\chi_{cJ}$ processes. Therefore,
it is informative to examine the polarized cross sections in
Table~\ref{Table-1} and Table~\ref{Table-2}.

For $J/\psi+\chi_{c0}$ production cross section, both the $(0,0)$
and $(1,0)$ helicity channels have comparable magnitude, either at
LO or at NLO, and the latter is even somewhat greater. This is quite
counterintuitive, diametrically contradicting what is expected from
the helicity selection rule. This might be viewed as a hint that the
$B$ factory energy may lie still far from the asymptotic scaling
regime.

For $e^+e^-\to J/\psi+\chi_{c1}$, the hierarchy among the cross
sections of three different helicity channels is also somewhat
abnormal. The contribution from the $(0,\pm 1)$ channel is far more
significant than the other two, and to a good approximation, one
only needs to retain this channel. This pattern still holds true
after incorporating the NLO correction for both settings of $\mu$.
It will be interesting for future Super $B$ experiments to verify
that the angular distribution of $J/\psi$ (or $\chi_{c1}$) is
predominantly of form $1+\cos^2\theta$. In passing, we also note
that the tiny LO contribution from the $(\pm 1,0)$ state may be
attributable to the accidental suppression factor with respect to
the helicity selection rule. After including the NLO correction,
$\sigma_{(1,0)}$ even becomes negative. This can be easily
understood by inspecting (\ref{asymp:NLO:K110}), since the NLO
helicity amplitude is $1/r$ enhanced relative to the LO one, and
with opposite sign. However, in practice this polarized cross
section is too small to be measured.

For $e^+e^-\to J/\psi+\chi_{c2}$, the hierarchy among the cross
sections of five different helicity channels roughly obeys the
helicity selection rule, except the contribution from the $(0,\pm
1)$ channel is much smaller than the $(\pm 1,0)$ channel. From
Table~\ref{Table-1} and Table~\ref{Table-2}, one sees that the bulk
of the cross sections comes from only two helicity states, {\it
i.e.} $(0,0)$ and $(\pm 1,0)$. For $\mu = \sqrt{s}/2$, the NLO
correction has small impact on both helicity states; for $\mu = 2
m_c$, the NLO corrections push down both polarized cross sections to
some extent. It would be interesting for the future high-statistics
experiment to observe this production process, and test whether the
produced $\chi_{c2}$ are predominantly longitudinally-polarized.

\section{Summary}
\label{summary}
\setcounter{equation}{0}

In this work we have computed the complete NLO perturbative
corrections to $e^+e^-\to J/\psi+\chi_{cJ}$ ($J=0,1,2$) within the
NRQCD factorization framework. The NLO NRQCD short-distance
coefficients are inferred by directly extracting the contribution
from the {\it hard} loop-momentum region. We have calculated the NLO
corrections to each of the 10 independent helicity amplitudes. We
have made a detailed analysis for both the polarized and unpolarized
cross sections and compared with the measurements at $B$ factory.

We find a significant positive NLO perturbative correction to
$e^+e^-\to J/\psi+\chi_{c0}$, which helps to bring the predicted
cross section in agreement with the $B$ factory measurements. In
contrast, our NLO predictions to $\psi'+\chi_{c0}$ cross section is
still significantly below the central value of the \textsc{Belle}
measurement. It is still too early to draw any solid conclusion
about whether NRQCD factorization fails for this channel or not. We
perhaps need to wait until \textsc{BaBar} collaboration carries out
an independent measurement for this process.

The impact of NLO corrections to $e^+e^-\to J/\psi+\chi_{c1,2}$
seems to be rather mild, even with their signs uncertain. Notice the
predicted cross sections for these processes are about 5 or 6 times
smaller than that for $e^+e^-\to J/\psi+\chi_{c0}$. Hopefully the
future Super $B$ factory, with much higher luminosity, will
eventually observe these two channels.

Our studies of polarized cross sections reveal that the bulk of the
total cross section comes from the $(0,\pm 1)$ helicity state for
$e^+e^-\to J/\psi+\chi_{c1}$, and from $(0,0)$ and $(\pm 1,0)$
helicity states for $e^+e^-\to J/\psi+\chi_{c2}$. It will be
interesting for the future Super $B$ experiments to test these
polarization patterns.

On the theoretical side, we have worked out the explicit asymptotic
expressions of all the ten NLO helicity amplitudes for the
$e^+e^-\to J/\psi+\chi_{cJ}$  ($J=0,1,2$) processes. We confirm that
helicity selection rule is modified logarithmically at NLO in
$\alpha_s$. The pattern we recognize in these NLO asymptotic
expressions lends further support for the speculation made in
Ref.~\cite{Jia:2010fw}: The hard exclusive processes involving
double-quarkonium at leading twist can only host the single
collinear logarithm $\ln{s/m_c^2}$ at one-loop, while those
beginning with the higher twist contributions are always plagued
with double logarithms of form $\ln^2{s/m_c^2}$.

It is of some theoretical interest to reproduce the asymptotic
expressions for the helicity-conserving channels such as $e^+e^-\to
J/\psi(0)+{\chi_{c0,2}}(0)$ in the light-cone approach. This should
be definitely feasible, which is guaranteed by the leading-twist
factorization theorem. Nevertheless, it is much more challenging for
the light-cone approach to reproduce, and resum, those
process-dependent double logarithms associated with the
helicity-suppressed channels.

{\noindent \it Note added.} After the calculation was finished and
while we were preparing the draft, a related work appeared in arXiv
recently~\cite{Wang:2011qg}, which also investigated the ${\cal
O}(\alpha_s)$ correction to the processes $e^+e^-\to
J/\psi+\chi_{cJ}$ ($J=0,1,2$) within NRQCD factorization approach.
Once taking the same input parameters as theirs, and summing up the
contributions from all the helicity states, our results agree with
Ref.~\cite{Wang:2011qg} on the numerical sizes of the NLO
corrections to the unpolarized cross sections for each $J=0,1,2$.

\begin{acknowledgments}
We are grateful to Loretta Robinette for valuable help on
proofreading.
This research was supported in part by the National Natural Science
Foundation of China under Grant No.~10875130, 10935012, by China
Postdoctoral Science Foundation, and by the U.~S. Department of
Energy under Grant No.~DE-FG02-93-ER40762.
\end{acknowledgments}


\begin{thebibliography}{99}

\bibitem{Lepage:1980fj}
  G.~P.~Lepage and S.~J.~Brodsky,
  Phys.\ Rev.\  D {\bf 22}, 2157 (1980).

\bibitem{Chernyak:1983ej}
  V.~L.~Chernyak and A.~R.~Zhitnitsky,
  Phys.\ Rept.\  {\bf 112}, 173 (1984).

\bibitem{Beneke:1999br}
  M.~Beneke, G.~Buchalla, M.~Neubert and C.~T.~Sachrajda,
  Phys.\ Rev.\ Lett.\  {\bf 83}, 1914 (1999)
  [arXiv:hep-ph/9905312].

\bibitem{Beneke:2000ry}
  M.~Beneke, G.~Buchalla, M.~Neubert and C.~T.~Sachrajda,
  Nucl.\ Phys.\  B {\bf 591}, 313 (2000)
  [arXiv:hep-ph/0006124].

\bibitem{Abe:2002rb}
  K.~Abe {\it et al.}  [\textsc{Belle} Collaboration],
  Phys.\ Rev.\ Lett.\  {\bf 89}, 142001 (2002).

\bibitem{Aubert:2005tj}
  B.~Aubert {\it et al.}  [BABAR Collaboration],
  Phys.\ Rev.\  D {\bf 72}, 031101 (2005)
  [arXiv:hep-ex/0506062].

\bibitem{Bodwin:1994jh}
  G.~T.~Bodwin, E.~Braaten and G.~P.~Lepage,
  Phys.\ Rev.\  D {\bf 51}, 1125 (1995)
  [Erratum-ibid.\  D {\bf 55}, 5853 (1997)]
  [arXiv:hep-ph/9407339].


\bibitem{Braaten:2002fi}
  E.~Braaten and J.~Lee,
  Phys.\ Rev.\  D {\bf 67}, 054007 (2003)
  [Erratum-ibid.\  D {\bf 72}, 099901 (2005)]
  [arXiv:hep-ph/0211085].

\bibitem{Liu:2002wq}
  K.~Y.~Liu, Z.~G.~He and K.~T.~Chao,
  Phys.\ Lett.\  B {\bf 557}, 45 (2003)
  [arXiv:hep-ph/0211181].

\bibitem{Hagiwara:2003cw}
  K.~Hagiwara, E.~Kou and C.~F.~Qiao,
  Phys.\ Lett.\  B {\bf 570}, 39 (2003)
  [arXiv:hep-ph/0305102].

\bibitem{Zhang:2005cha}
  Y.~J.~Zhang, Y.~j.~Gao and K.~T.~Chao,
  Phys.\ Rev.\ Lett.\  {\bf 96}, 092001 (2006)
  [arXiv:hep-ph/0506076].

\bibitem{Gong:2007db}
  B.~Gong and J.~X.~Wang,
  Phys.\ Rev.\  D {\bf 77}, 054028 (2008)
  [arXiv:0712.4220 [hep-ph]].

\bibitem{Ma:2004qf}
  J.~P.~Ma and Z.~G.~Si,
  Phys.\ Rev.\  D {\bf 70}, 074007 (2004)
  [arXiv:hep-ph/0405111].

\bibitem{Bondar:2004sv}
  A.~E.~Bondar and V.~L.~Chernyak,
  Phys.\ Lett.\  B {\bf 612}, 215 (2005).

\bibitem{Braguta:2008tg}
  V.~V.~Braguta,
  arXiv:0811.2640 [hep-ph].

\bibitem{Zhang:2008gp}
  Y.~J.~Zhang, Y.~Q.~Ma and K.~T.~Chao,
  Phys.\ Rev.\  D {\bf 78} (2008) 054006
  [arXiv:0802.3655 [hep-ph]].

\bibitem{:2009nj}
  P.~Pakhlov {\it et al.} [ Belle Collaboration ],
  Phys.\ Rev.\  {\bf D79}, 071101 (2009).
  [arXiv:0901.2775 [hep-ex]].

\bibitem{Jia:2010fw}
  Y.~Jia, J.~X.~Wang and D.~Yang,
  arXiv:1012.6007 [hep-ph].

\bibitem{Jacob:1959at}
  M.~Jacob and G.~C.~Wick,
  Annals Phys.\  {\bf 7}, 404 (1959)
  [Annals Phys.\  {\bf 281}, 774 (2000)].

\bibitem{Haber:1994pe}
  H.~E.~Haber,
  arXiv:hep-ph/9405376.

\bibitem{Brodsky:1981kj}
  S.~J.~Brodsky and G.~P.~Lepage,
  Phys.\ Rev.\  D {\bf 24}, 2848 (1981).

\bibitem{Kuhn:1979bb}
  J.~H.~Kuhn, J.~Kaplan and E.~G.~O.~Safiani,
  Nucl.\ Phys.\  B {\bf 157}, 125 (1979).

\bibitem{Bodwin:2002hg}
  G.~T.~Bodwin and A.~Petrelli,
  Phys.\ Rev.\  D {\bf 66}, 094011 (2002)
  [arXiv:hep-ph/0205210].

\bibitem{Kublbeck:1990}
  J.~Kublbeck, M.~bohm and A,~Denner,
  Comput.\ Phys.\ Commun.\  {\bf 60}, 165 (1990).

\bibitem{Hahn:2000kx}
  T.~Hahn,
  Comput.\ Phys.\ Commun.\  {\bf 140} (2001) 418
  [arXiv:hep-ph/0012260].



\bibitem{Metig:1991}
  R.~Mertig, M.~Bohm, and A.~Denner,
  Comput.\ Phys.\ Commun. {\bf 64}, 345 (1991)


\bibitem{Beneke:1997zp}
  M.~Beneke and V.~A.~Smirnov,
  Nucl.\ Phys.\  B {\bf 522}, 321 (1998)
  [arXiv:hep-ph/9711391].


\bibitem{Bodwin:2008nf}
  G.~T.~Bodwin, X.~Garcia i Tormo and J.~Lee,
  Phys.\ Rev.\ Lett.\  {\bf 101}, 102002 (2008)
  [arXiv:0805.3876 [hep-ph]].

\bibitem{Song:2002mh}
  Z.~z.~Song and K.~T.~Chao,
  Phys.\ Lett.\  B {\bf 568}, 127 (2003)
  [arXiv:hep-ph/0206253].

\bibitem{Song:2003yc}
  Z.~Z.~Song, C.~Meng, Y.~J.~Gao and K.~T.~Chao,
  Phys.\ Rev.\  D {\bf 69}, 054009 (2004)
  [arXiv:hep-ph/0309105].
\bibitem{Beneke:2008pi}
  M.~Beneke and L.~Vernazza,
  Nucl.\ Phys.\  B {\bf 811}, 155 (2009)
  [arXiv:0810.3575 [hep-ph]].

\bibitem{Smirnov:2008iw}
  A.~V.~Smirnov,
  JHEP {\bf 0810}, 107 (2008).
  [arXiv:0807.3243 [hep-ph]].

\bibitem{Hahn:1998yk}
  T.~Hahn and M.~Perez-Victoria,
  Comput.\ Phys.\ Commun.\  {\bf 118}, 153 (1999).

\bibitem{Jia:2008ep}
  Y.~Jia and D.~Yang,
  Nucl.\ Phys.\  B {\bf 814} (2009) 217
  [arXiv:0812.1965 [hep-ph]].


\bibitem{Jia:2007hy}
  Y.~Jia,
  Phys.\ Rev.\  D {\bf 76} (2007) 074007
  [arXiv:0706.3685 [hep-ph]].

\bibitem{Gong:2008ue}
  B.~Gong, Y.~Jia and J.~X.~Wang,
  Phys.\ Lett.\  B {\bf 670} (2009) 350
  [arXiv:0808.1034 [hep-ph]].

\bibitem{Lepage:1979zb}
  G.~P.~Lepage and S.~J.~Brodsky,
  Phys.\ Lett.\  B {\bf 87}, 359 (1979).


\bibitem{Efremov:1979qk}
  A.~V.~Efremov and A.~V.~Radyushkin,
  Phys.\ Lett.\  B {\bf 94} (1980) 245.

\bibitem{Bodwin:2007ga}
  G.~T.~Bodwin, J.~Lee, C.~Yu,
  Phys.\ Rev.\  {\bf D77}, 094018 (2008).
  [arXiv:0710.0995 [hep-ph]].

\bibitem{Eichten:1995}
  E.~J.~Eichten and C.~Quigg,
  Phys.\ Rev.\  D {\bf 52}, 1726 (1995)
  [arXiv:hep-ph/9503356].

\bibitem{Buchmuller:1981}
  W.~Buchm$\ddot{\mbox{u}}$ller and S.-H.~H.~Tye,
  Phys.\ Rev.\  D {\bf 24}, 132 (1981)
  [arXiv:hep-ph/9503356].

\bibitem{Abe:2004ww}
  K.~Abe {\it et al.}  [Belle Collaboration],
  Phys.\ Rev.\  D {\bf 70}, 071102 (2004)
  [arXiv:hep-ex/0407009].


\bibitem{Wang:2011qg}
  K.~Wang, Y.~-Q.~Ma, K.~-T.~Chao,
  Phys.\ Rev.\  {\bf D84}, 034022 (2011).
  [arXiv:1107.2646 [hep-ph]].

\end{thebibliography}
\end{document}


\preprint{
\vbox{
\halign{&#\hfil\cr
    & January 2013 \cr
    & \cr}}
}

\title{Erratum: $\bm{\mathcal O}\bm{(}\bfalpha_{\bm s}\bm{)}$ corrections to
$\bm{J}\bm{/}\bfpsi \bm{+}\bfchi_{\bm{c}\bm{J}}$ production at
$\bm{B}$ factories \\
{[JHEP {\bf 1110}, 141 (2011)]}}

\author{Hai-Rong Dong\footnote{E-mail: donghr@ihep.ac.cn}}
\affiliation{Institute of High Energy Physics, Chinese Academy of
Sciences, Beijing 100049, China\vspace{0.2cm}}

\author{Feng Feng\footnote{E-mail: fengf@ihep.ac.cn}}
\affiliation{Center for High Energy Physics, Peking University,
Beijing 100871, China\vspace{0.2cm}}

\author{Yu Jia\footnote{E-mail: jiay@ihep.ac.cn}}
\affiliation{Institute of High Energy Physics, Chinese Academy of
Sciences, Beijing 100049, China\vspace{0.2cm}}
\affiliation{Theoretical Physics Center for Science Facilities,
Chinese Academy of Sciences, Beijing 100049, China\vspace{0.2cm}}

\maketitle

\renewcommand{\theequation}{\thesection.\arabic{equation}}
\renewcommand{\thesection}{\arabic{section}}


In Ref.~\cite{Dong:2011fb}, we study the ${\cal O}(\alpha_s)$
corrections to $e^+e^-\to J/\psi(\psi')+\chi_{cJ}$ ($J=0,1,2$) in
NRQCD factorization approach. These next-to-leading order (NLO)
perturbative corrections are calculated at the level of helicity
amplitude. There are some errors in equations (3.3), (3.4), (4.5),
(4.6) and (4.7). Fortunately, these incorrect expressions have not
been used in our computer program, therefore all the figures and
tables in Ref.~\cite{Dong:2011fb} need not be modified.

There is one typo in the tree-level NRQCD short-distance coefficient
associated with the helicity state $\gamma^* \rightarrow J/\psi(\pm
1)+\chi_{c2}(0)$, as given in (3.3c). The corrected coefficients
$c^{J}_{\lambda_1,\lambda_2}(r)$ ($J=0,1,2$) now read
\setcounter{section}{3}
\setcounter{equation}{2}
\begin{subequations}
\bqa
& & c^0_{0,0}(r)= 1+ 10r - 12r^2\qquad c^0_{1,0}(r)= 9-14r,
\\
& & c^1_{0,1}(r)=-\sqrt{6}(2-7 r)\qquad c^1_{1,0}(r)=-\sqrt{6}
r\qquad c^1_{1,1}(r)= -2\sqrt{6}(1-3r),
\label{c:function:LO:chic1}
\\
& & c^2_{0,0}(r)=\sqrt{2}(1-2r-12r^2)\qquad
c^2_{0,1}(r)=\sqrt{6}(1-5r)\qquad c^2_{1,0}(r)= \sqrt{2}(3-11r),
\nn\\
& & c^2_{1,1}(r) = 2\sqrt{6}(1-3r)\qquad c^2_{1,2}(r)= 2\sqrt{3}.
\eqa
\end{subequations}
Note that $c^2_{1,0}(r)$ was mistyped as $\sqrt{2}(11-3r)$ in
Ref.~\cite{Dong:2011fb}. To be compatible with the phase convention
of the helicity amplitude projector used in Ref.~\cite{Xu:2012uh},
we also include a minus sign for some entries in (3.3). These
changes obviously do not affect our numerical predictions for the
polarized cross sections.

For the expression that relates the coefficients
$c^{J}_{\lambda_1,\lambda_2}(r)$ and the leading-order polarized
cross section, as given in (3.4), we have missed a factor
$r^{1+|\lambda_1+\lambda_2|}$ in the right side. The correct formula
should read
\bqa
& & \sigma^{(0)}[J/\psi(\lambda_1) + \chi_{cJ}(\lambda_2)] = {32\pi
e_c^2 \alpha^2 C_F^2 \alpha_s^2 \over 3 s^2 m_c^6} \left({|{\bf
P}|\over \sqrt{s}}\right) R^2_{J/\psi}(0) R'^2_{\chi_{cJ}}(0) r^{1+|\lambda_1+\lambda_2|} \left|
c^J_{\lambda_1,\lambda_2}(r)\right|^2.
\label{pol:cross:section:LO:alphas}
\eqa

We also made an error on the overall normalization in the asymptotic
expressions of all the ten reduced NLO helicity amplitudes,
$K^{J}_{\lambda_1,\lambda_2}$ ($J=0,1,2$) in
Ref.~\cite{Dong:2011fb}. All the expressions (except $K^{2}_{1,0}$)
should be multiplied by an overall factor of 4. Since our computer
code implemented the correct asymptotic expressions, Figures~2, 3, 4
in Ref.~\cite{Dong:2011fb} remain intact.

\setcounter{section}{4}
\setcounter{equation}{4}
The asymptotic expressions of two reduced NLO helicity amplitudes
for $J/\psi+\chi_{c0}$ should read
\begin{subequations}
\bqa
& & K^0_{0,0} \left( r,{\mu^2 \over s} \right)_{\rm asym} = -{1\over
3} (4-\ln2) \ln r + {\beta_0 \over 4} \bigg( \ln {4 \mu^2\over s} +
{8\over 3}\bigg)
\label{asymp:NLO:K000}
\nn \\
&& - \frac{1}{18} ( 46 + \pi^2 - 40\ln2 + 33 \ln^2 2) + {i\pi\over
4} \bigg(\beta_0 -{16\over 3}+ {4\over 3}\ln2 \bigg),
\\
& &  K^0_{1,0}\left( r,{\mu^2 \over s} \right)_{\rm asym} = {1\over
3} \ln^2 r - {1 \over 108}(139-104\ln2)\ln r + {\beta_0 \over 4}
\bigg( \ln{4 \mu^2 \over s} + {17\over 9} \bigg)
\nn \\
&& - {1\over 54}\bigg( 161 + {8\pi^2\over 3}- {495\over 2}\ln2 + 100
\ln^2 2 \bigg) + {i\pi \over 4} \bigg( {8\over 3}\ln r + \beta_0 -
{1\over 27}(139 -104\ln 2)\bigg).
\nn\\
\eqa
\label{K:asymp:NLO:Jpsi:chic0}
\end{subequations}

The asymptotic expressions of the three reduced NLO helicity
amplitudes for $J/\psi+\chi_{c1}$ should read
\begin{subequations}
\bqa
& & r\,K^1_{1,0}\left( r,{\mu^2 \over s} \right)_{\rm asym} =
-{1\over 12} \Bigg\{ 5 \ln^2r + (7-2\ln2)\ln r -19 + 2\pi^2 + 75\ln
2- 21 \ln^2 2
\label{asymp:NLO:K110}
\nn \\
&& + i\pi ( 10\ln r + 7 - 2\ln2 )\Bigg\},
\\
& & K^1_{0,1}\left( r,{\mu^2 \over s} \right)_{\rm asym} = {1\over
24}\Bigg\{ \frac{25}{2} \ln^2 r - (46-99\ln2)\ln r + 6\beta_0 \bigg(
\ln{4\mu^2\over s}  + {13\over 6}\bigg)
\nn \\
&& - {1\over 6} (616 + 74\pi^2 - 1696\ln2 + 303 \ln^22 ) +
i\pi(25\ln r + 6 \beta_0 -46 + 99\ln2 )\Bigg\},
\\
& & K^1_{1,1} \left( r,{\mu^2 \over s} \right)_{\rm asym} = {1\over
24}\Bigg\{ 10 \ln^2r + 2(1+17\ln2)\ln r + 6\beta_0 \bigg(
\ln{4\mu^2\over s}  + {13\over 6}\bigg)
\nn \\
&& - \frac{1}{3}( 266 + 7\pi^2 - 128\ln 2 + 147 \ln^2 2) + 2 i\pi(
10\ln r + 3\beta_0 + 1+ 17\ln2) \Bigg\}.
\eqa
\label{K:asymp:NLO:Jpsi:chic1}
\end{subequations}

The asymptotic expressions of the five reduced NLO helicity
amplitudes for $J/\psi+\chi_{c2}$ should read
\begin{subequations}
\bqa
& & K^2_{0,0}\left( r,{\mu^2 \over s} \right)_{\rm asym} = -{1\over
3}(4-\ln2) \ln r + {\beta_0 \over 4} \bigg(\ln{4\mu^2 \over s} +
{8\over 3} \bigg)
\nn \\
&& -\frac{1}{18} ( 64 + \pi^2 + 104\ln 2 + 33 \ln^22) +
 \frac{i\pi}{4} \bigg( \beta_0 - {10 \over 3} + {4\over 3}\ln2 \bigg),
\label{asymp:NLO:K200}
\\
& & K^2_{0,1} \left( r,{\mu^2 \over s} \right)_{\rm asym} = {1\over
12} \Bigg\{ \frac{13}{2}\ln^2 r - (22-43\ln2) \ln r + 3 \beta_0
 \bigg(\ln{4\mu^2 \over s} +
{8\over 3} \bigg)
\nn \\
&& - \frac{1}{6} ( 284 + 30\pi^2 - 380\ln2 + 159 \ln^2 2) + i\pi
(13\ln r + 3 \beta_0 - 14 + 43\ln2 )\Bigg\},
\\
& & K^2_{1,0}\left( r,{\mu^2 \over s} \right)_{\rm asym} = {1\over
6} \Bigg\{2\ln^2 r +\frac{1}{6}(5+8\ln2) \ln r + {3\over 2}\beta_0
\bigg(
 \ln {4\mu^2\over s} + {7\over 3} \bigg)
\nn\\
&& -\frac{1}{18} ( 291-8\pi^2 + 171 \ln 2 + 312\ln^2 2 ) + i\pi
\bigg( 4\ln r + {3\over 2} \beta_0 + {11\over 6} + {4 \over 3} \ln2
\bigg)\Bigg\},
\\
& & K^2_{1,1} \left( r,{\mu^2 \over s} \right)_{\rm asym}  = {1\over
24} \Bigg\{ 4\ln^2 r - (46-62\ln2) \ln r + 6 \beta_0 \bigg(
\ln{4\mu^2\over s} + {13\over 6} \bigg )
\nn\\
&& - \frac{1}{3}(274 + 27\pi^2 - 316\ln 2 + 9\ln^2 2) + i\pi (8\ln r
+ 6 \beta_0 -46 +62\ln 2)\Bigg\},
\\
& & K^2_{1,2} \left( r,{\mu^2 \over s} \right)_{\rm asym} = -{1\over
4} \Bigg\{ 2\ln^2 r + \frac{2}{3}(1+13\ln2) \ln r - \beta_0 \bigg(
\ln{4\mu^2\over s} + {5\over 3} \bigg)
\nn\\
&& + \frac{1}{9} ( -7\pi^2 + 140 - 104\ln 2 + 237\ln^2 2)- i\pi
\bigg(\beta_0 + 3 - {26\over 3} \ln2 \bigg)\Bigg\}.
\eqa
\label{K:asymp:NLO:Jpsi:chic2}
\end{subequations}
Note that the overall factor in (4.7c) was mistakenly typed as
$-{1\over 88}$ in Ref.~\cite{Dong:2011fb}.
In the current formulas, the renormalization scale $\mu$ enters the
$K^J_{\lambda_1,\lambda_2}$ functions in the form of ${\beta_0 \over
4} \ln{4\mu^2 \over s}$ unanimously, which stems from the
renormalization group invariance of each NLO polarized cross section
for $e^+e^-\to J/\psi+\chi_{c0,1,2}$.